\documentclass[prl,reprint,superscriptaddress,
amsmath,amssymb,aps,]{revtex4-2}
\usepackage{array}
\usepackage{amsmath}
\usepackage{graphicx}
\usepackage{dcolumn}
\usepackage{bm}
\usepackage{float}
\usepackage{hyperref}
\usepackage{physics}
\hypersetup{colorlinks,linkcolor={blue},citecolor={blue},urlcolor={blue}}  
\usepackage{xtab,afterpage,longtable}
\usepackage[utf8]{inputenc}
\DeclareUnicodeCharacter{03C3}{\ensuremath{\sigma}}
\makeatletter
\def\LT@LR@e{\LTleft\z@   \LTright\z@}%
\makeatother

\begin{document}

\preprint{APS/123-QED}

\title{Disorder Enhanced Thermalization in Interacting Many-Particle System}
\author{Chakradhar Rangi}
\email{crangi1@lsu.edu}
\affiliation{Department of Physics and Astronomy, Louisiana State University, Baton Rouge, LA 70803, USA}
\author{Herbert F Fotso}
\affiliation{Department of Physics, University at Buffalo SUNY, Buffalo, New York 14260, USA}
\author{Hanna Terletska}
\affiliation{Department of Physics and Astronomy, Middle Tennessee State University, Murfreesboro, TN 37132, USA}
\author{Juana Moreno}
\affiliation{Department of Physics and Astronomy, Louisiana State University, Baton Rouge, LA 70803, USA}
\affiliation{Center for Computation and Technology, Louisiana State University, Baton Rouge, LA 70803, USA}
\author{Ka-Ming Tam}
\affiliation{Department of Physics and Astronomy, Louisiana State University, Baton Rouge, LA 70803, USA}
\affiliation{Center for Computation and Technology, Louisiana State University, Baton Rouge, LA 70803, USA}
\date{\today}
\begin{abstract}

We introduce an extension of the non-equilibrium dynamical mean field theory to incorporate the effects of static random disorder in the dynamics of a many-particle system by integrating out different disorder configurations resulting in an effective time-dependent density-density interaction. We use this method to study the non-equilibrium  transient dynamics of a system described by the Fermi Anderson-Hubbard model following an interaction and disorder quench. The method recovers the solution of the disorder-free case for which the system exhibits qualitatively distinct dynamical behaviors in the weak-coupling (prethermalization) and strong-coupling regimes (collapse-and-revival oscillations). However, we find that weak random disorder promotes thermalization. In the weak coupling regime, the jump in the quasiparticle weight in the prethermal regime is suppressed by random disorder while in the strong-coupling regime, random disorder reduces the amplitude of the quasiparticle weight oscillations. These results highlight the importance of disorder in the dynamics of realistic many-particle systems.

\end{abstract}\maketitle

Driven by theoretical and experimental advances, the nonequilibrium properties of correlated many-particle systems have recently  been a subject of increased interest \cite{Greiner2002,Eckstein2009,Wall2011,QuantumQuenchSYK,ReviewNEDMFT,Lantz2017}.
Within this field, thermalization after interaction quenches has been a major focus \cite{QuenchBoseHubbard,Peter_HeisenbergChains,ThermalizationSYK,IntQuenchHubbard,EcksteinPRLThermalization_2009,Peter_Typ_Prethermalization2019,Karsten_PRX,DPT_review_2022}. To probe thermalization in interacting fermionic systems, a simple quantity to monitor is the time dependence of the distribution function, particularly near the Fermi level. Studies of interaction quenches in the half-filled Hubbard model typically reveal two distinct dynamical regimes \cite{IntQuenchHubbard,PrethermalGGEKollar_2011,EcksteinPRLThermalization_2009,DPT_review_2022,EcksteinPRBInteractionquench_2010,TDMFT_Quench2010,TDMFT_QuenchPRB}. In the weak coupling regime, the initial jump in the distribution function immediately after the interaction quench saturates to a finite value that persists for a long time.  On the contrary, in the strong coupling regime, the jump in the distribution function decays over time and exhibits damped oscillations around a finite value. This qualitative difference in behavior suggests a dynamical phase transition \cite{DPT_review} separating these two thermalization regimes in the half-filled Hubbard model \cite{DPT_review_2022}. 

Partly due to the proposal of many-body localization, extensive efforts have been devoted to studying the effect of strong disorder in one dimension, seeking evidence for this phenomenon \cite{MBL_RMP,Alet_Laflorencie_2018,Nandkishore_Huse_2015}. In contrast, there have been a limited number of studies on the effect of thermalization with weak random disorder beyond one dimension. The presence of some form of weak disorder is often inevitable in condensed matter experiments \cite{Vojta_disorder_review,RevModPhys.46.465}. Characterizing its effect, particularly the effect of weak random disorder, is thus crucial not only for understanding localization expected in strong disorder but also for interpreting experimental results, where weak disorder is almost always present.

Tackling disorder is inherently challenging for computational studies. One commonly used approach is the coherent potential approximation (CPA) \cite{SovenP_CPA_Apr1967,Velicky_CPA_Aug1969,Kirkpatrick_CPA_Apr1970,Shiba_CPA_1971,Yonezawa1CPA_1973}. On the other hand, dynamical mean field theory (DMFT) \cite{WMetzner_DMFT_Jan1989,Georges_DMFTReview,Kotliar_DMFTReview,Freericks_DMFTReview}
and its cluster extensions \cite{Thomas_ClusterDMFTReview,Hettler_ClusterDMFT_Sep1998,Hettler_ClusterDMFT_May2000,Kotliar_CellularDMFT_Oct2001,Jarrell_ClusterDMFT_Oct2001,TMDCA_review,realspaceTMT,Fotso2012,Fotso_2022} have proven to be successful methods for handling interaction, providing insights into the physics of strongly correlated systems. The combination of CPA and DMFT has facilitated numerous explorations into the intricate interplay between disorder and electronic correlations within condensed matter systems \cite{Janis_CPADMFT_1992Dec,Kuchinskii_CPADMFT_2010,Semmler_CPADMFT_2011Sept,Miranda_CPADMFT_2012June,Weh_CPADMFT_2021Jul,Ekuma_etal_2015}. 

The advancement of experimental techniques on various pump-and-probe experiments allow the study of time dynamics of correlated systems \cite{dong2017pump,grunbein2020illumination,fischer2016invited}. They present an intriguing avenue of exploration for understanding the interplay between disorder and electronic correlations away from equilibrium. On the other hand, computational methods have also been actively developing \cite{ReviewNEDMFT}. Notably, recent endeavors have extended the CPA and DMFT combination to the non-equilibrium framework \cite{Dohner_NEDMFTCPA,YanJiawei_NEDMFTCPA,Dohner_Thermalization_Oct2023,RangiOTOC_2024}.
In this work, we propose an alternate approach to study non-equilibrium dynamics in disordered systems in the non-equilibrium DMFT formalism by integrating out the disorder exactly within the Schwinger-Keldysh framework \cite{Horbach_Schon_1993,Kamenev_Andreev_1999,CChamon1999Jul}. This strategy leverages the normalization property of the Schwinger partition function, circumventing the need for explicit disorder averaging over numerous realizations and thus offering significant computational advantages. By using the method developed, we are able to study the effect of disorder on the thermalization. These studies reveal that disorder facilitates thermalization of the many-particle system. 

\textit{Model} -- We consider a system described by the single-band Anderson-Hubbard model. It is a simple model that captures the interplay of electron-electron interaction and random disorder. The Hamiltonian is given by:
    \begin{gather}
        H = -\sum_{\expval{ij}\sigma} t_{ij}\Bigl(c^\dagger_{i\sigma} c_{j\sigma} + \text{h.c.}\Bigr) +  \sum_{i\sigma}\Bigl(V_i(t) - \mu\Bigr)n_{i\sigma}\nonumber\\+ 
        U(t)\sum_i  n_{i\uparrow}n_{i\downarrow},
        \label{eqn:Anderson_Hubbard}
    \end{gather}
    \vspace{-0.02cm}
    where $c^\dagger_{j\sigma}$($c_{j\sigma}$) represents the electron creation (annihilation) operator at site $j$ and spin $\sigma$ and $n_{j\sigma} \equiv c^\dagger_{j\sigma}c_{j\sigma}$ is the electron number operator. The first term represents the kinetic energy arising from the nearest neighbour hopping with $t_{ij} = t_{hop}$ denoting the hopping amplitude. We set $t_{hop}=0.25$ and it serves as the energy scale.  The second term contains the local on-site disorder potential $V_j$ and the chemical potential $\mu$. Here, we assume that $V_j$ is randomly distributed according to a Gaussian probability distribution without correlation among different sites. The last term represents the electron-electron interaction energy with $U$ denoting the interaction strength.
    
\textit{Nonequilibrium formalism} --
    Our nonequilibrium theory is formulated on the threefold Schwinger-Keldysh contour $\mathcal{C}$. The contour extends from an initial time $t_0$ to a maximum time of interest $t_f$ along the real time axis ($\mathcal{C}_1$), returning back to $t_0$ ($\mathcal{C}_2$), and then to $t=t_0-i\beta$ on the imaginary time axis ($\mathcal{C}_3$). The system is in equilibrium initially at $t_0$ with inverse temperature, $\beta$, and subsequently driven out of equilibrium by external fields or by an interaction quench. Such an evolution of the vacuum state from $t_0$ to $t_f$ and back to $t_0$ implies that the nonequilibrium partition function $\mathcal{Z}$, defined on the real-time contour is normalized to unity: $\mathcal{Z}^{\mathcal{C}_1\cup\mathcal{C}_2}=1$. 
    The partition function $\mathcal{Z}^{\mathcal{C}}$, defined on the contour $\mathcal{C}$ has an important factorization property \cite{landsman_1987,Horbach_Schon_1993,CChamon1999Jul}, namely, $\mathcal{Z}^{\mathcal{C}} = \mathcal{Z}^{\mathcal{C}_1\cup\mathcal{C}_2 } \mathcal{Z}^{\mathcal{C}_3}.$
    The computation of the real-time correlation functions, only depend on the factor $\mathcal{Z}^{\mathcal{C}_1\cup\mathcal{C}_2 }$. The other factor, $\mathcal{Z}^{\mathcal{C}_3}$, acts merely as a multiplicative constant in this computation, which entirely drops out. Thus, we can still work with a partition function normalized to unity at finite temperatures to compute the real-time correlation functions. A significant advantage follows that the partition function is automatically normalized $\mathcal{Z}=1$ for any realization of the disorder potential. This eliminates the denominator problem that typically arises when computing disorder-averaged real-time correlation functions. 
    
    Let us now consider the nonequilibrium partition function $\mathcal{Z}_V$ of the Anderson-Hubbard Hamiltonian (Eq. \eqref{eqn:Anderson_Hubbard}) defined on $\mathcal{C}$ for a specific realization of the disorder potential V: $
        \mathcal{Z}_V = \int_\mathcal{C} \mathcal{D}\bigl[\bar{\psi},\psi\bigr] e^{i\mathcal{S}_\text{lat}(\bar{\psi},\psi)}$,
    where $(\bar{\psi},\psi)$ denote the Grassmann fields. The lattice action $\mathcal{S}_\text{lat}$ is given by:
    \begin{gather}
    \mathcal{S}_\text{lat} = \sum_{ij\sigma}\int_\mathcal{C} \dd{t} \dd{t'} \Bar{\psi}_{i \sigma}(t)\mathcal{G}^{-1}_{i j \sigma,V_i}(t,t')\psi_{j \sigma}(t') \nonumber \\ - \sum_i \int_\mathcal{C} \dd{t}  U(t)\Bar{\psi}_{i\uparrow}(t)\Bar{\psi}_{i\downarrow}(t)\psi_{i\downarrow}(t)\psi_{i\uparrow}(t).
\end{gather}
    Here, $\mathcal{G}^{-1}_{i j \sigma,V_i}(t,t')  \equiv
    \Big[ \Bigl(i\pdv{t} + \mu - V_i\Bigr) \delta_{i j} + t_{ij}\Big]\delta_c(t,t')$, denotes the non-interacting lattice Green's function. 
    Considering the random disorder with Gaussian distribution according to $P[V] \propto \exp{-\frac{1}{2W^2}\sum_i V^2_i}$ 
    with $W^2$ denoting the variance of the distribution. We can compute the disorder averaged partition function by integrating over $V$ using the property that the Schwinger-Keldysh formalism cancels the denominator as discussed before:
    $\mathcal{Z} = \int \mathcal{D}V P[V] \mathcal{Z}_V 
    \equiv \int_\mathcal{C} \mathcal{D}\bigl[\bar{\psi},\psi\bigr] e^{i\mathcal{S}_0(\bar{\psi},\psi)} e^{i\mathcal{S}_\text{dis}(\bar{\psi},\psi)} \label{averaged_part}$. Here $\mathcal{S}_0$ is the lattice action without the disorder potential term. We can identify the effective action after integrating the disorder as
    \begin{gather}
        \mathcal{S}_\text{dis}(\bar{\psi},\psi) = i\frac{W^2}{2} \times \nonumber \\ \sum_{i, \sigma,\sigma'} \int_{\mathcal{C}}\dd{t} \dd{t'} \Bar{\psi}_{i \sigma}(t)\psi_{i \sigma}(t) \Bar{\psi}_{i \sigma'}(t')\psi_{i \sigma'}(t').
        \label{eqn:eff_vertex}
    \end{gather}
    Thus, integrating out the Gaussian distributed disorder term generates an effective interaction vertex involving density-density interaction at different times with an interaction strength, $\Bar{U} \equiv i\frac{W^2}{2}$. This manipulation of disorder is formally exact. The exact disorder integration is limited to disorders with Gaussian distributions. Any other disorder distribution would not necessarily lead to exact integrations and would require some approximations to analytically integrate over disorder configurations. A general disorder distribution can in principle be studied by expanding the distribution function as the moments of Gaussian, and this will end up with effective high order interaction.
    
    To proceed further, we use the non-equilibrium DMFT approximation to map this complex problem of interacting electrons on a lattice into a single-site impurity problem in an effective dynamical bath \cite{Freericks_BlochOsc_Feb2008}. The key approximation is that the electronic self-energy is assumed to be local in space. This simplification is exact in the limit of infinite dimensions, it makes the problem significantly more tractable, allowing for numerical and semi-analytical solutions \cite{Georges_DMFTReview,ReviewNEDMFT}. The impurity action is given by: 
    \begin{gather}
        \mathcal{S}_{\text{imp}} = \sum_{\sigma}\int_\mathcal{C} \dd{t} \dd{t'} \Bar{\psi}_\sigma(t)\mathcal{G}^{-1}_{0,\sigma}(t,t')\psi_\sigma(t') \nonumber \\ -  \int_\mathcal{C} \dd{t} U(t)\Bar{\psi}_{\uparrow}(t)\Bar{\psi}_{\downarrow}(t)\psi_{\downarrow}(t)\psi_{\uparrow}(t)  \nonumber\\ + \Bar{U}\sum_{\sigma,\sigma'} \int_{\mathcal{C}}\dd{t} \dd{t'} \Bar{\psi}_\sigma(t)\psi_\sigma(t) \Bar{\psi}_{\sigma'}(t')\psi_{\sigma'}(t').
    \end{gather}
    Here, $\mathcal{G}^{-1}_{0,\sigma}(t,t')\equiv\Bigl(i\pdv{t} + \mu\Bigr)\delta_c(t,t') - \Lambda_\sigma(t,t')$, denotes the non-interacting impurity Green's function and $\Lambda_\sigma(t,t')$ denotes the hybridization function.
    One obtains the impurity self-energy, $\Sigma_{\sigma}(t,t')$, using an impurity solver and further computes the local Green's function $G_\sigma(t,t)$ using the Dyson equation. For our model on the Bethe lattice, the self-consistency condition reduces to $\Lambda_\sigma(t,t') = t_{hop}^2G_\sigma(t,t')$ \cite{ReviewNEDMFT,Georges_DMFTReview}. We use the perturbation theory to obtain the self-energy and the Green's function of the impurity site. The local interaction at the impurity site contains two interaction vertex, namely, the Hubbard $U$ and the disorder induced non-local in time effective vertex $\Bar{U}$. The bare perturbation (IPT) at the lowest order does not capture the expected physics of the interplay of disorder and interactions. We treat the disorder vertex at the first order perturbation diagram using the self-consistent perturbation theory (SPT) and the Hubbard $U$ using the second-order bare IPT. The impurity self-energy is given by:
    \begin{gather}
        \Sigma_\sigma(t,t') =  U(t)U(t')G^{\Bar{U}}_\sigma(t,t')G^{\Bar{U}}_\sigma(t,t')G^{\Bar{U}}_\sigma(t',t)\nonumber\\ 
        - i\Bar{U}(t)G^{\Bar{U}}_\sigma(t,t').
    \end{gather}
    Here, $G^{\Bar{U}}_\sigma(t,t')$ is self-consistently obtained using the diagrams involving only the disorder induced effective vertex. Notably, the second-order bare IPT reproduces the strong-coupling behavior at half-filling, capturing the Mott metal-insulator transition with accuracy comparable to QMC results \cite{WeakCoupling_Naoto}. However, the second order bare IPT impurity solver has known limitations, particularly in conserving energy at intermediate to longer timescales in non-equilibrium DMFT for strong coupling. Given our focus on short-time dynamics, especially the initial revival oscillation of the Fermi surface discontinuity in the strong coupling regime, as discussed later in the text, we believe IPT is suitable for our current investigation. 
    
    For practical calculation the Schwinger-Keldysh contour is discretized into a finite number of time steps along the entire contour. Under this approximation, the two time non-equilibrium Green's function can be represented in term of a finite size $N \times N$ matrix, where $N$ is the total number of time steps on the contour. The bare Green's function is given by $ G_{0}(t,t^{'}) = \left[(i\partial_t + \mu) \delta \right]^{-1}(t,t^{'})$. It can be obtained through the numerical implementation of Ref \cite{Freericks_BlochOsc_Feb2008}. Alternatively, more sophisticated numerical techniques, such as those discussed in Ref \cite{NESSi}, can be employed to achieve higher accuracy.

\begin{figure}[tbp]
        \centering
        \begin{minipage}{0.5\columnwidth}
            \centering
            \includegraphics[width=\columnwidth]{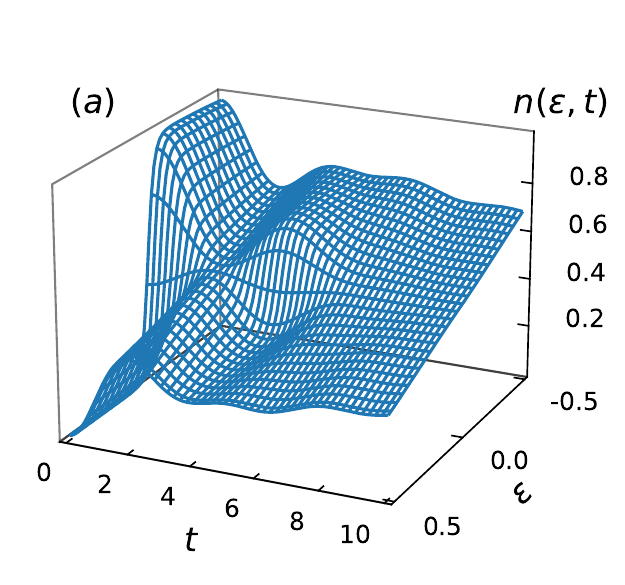} 
            \label{fig:Net_U1.5W0}
        \end{minipage}\hfill
        \begin{minipage}{0.5\columnwidth}
            \centering
            \includegraphics[width=\columnwidth]{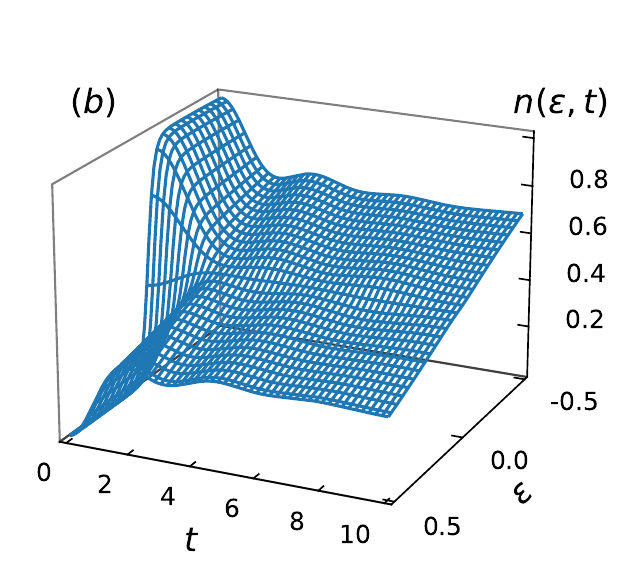} 
            \label{fig:Net_U1.5W0.25}
        \end{minipage}
        \caption{The momentum distribution function $n(\varepsilon,t)$ for an interaction quench from $U=0$ to $U=1.5$ for (a) clean system limit $\Omega=0.0$ (b) weak disorder $\Omega=0.5$.}
        \label{fig:MomDistributionFunc}
    \end{figure}

\textit{Results} -- We proceed by investigating the non-equilibrium evolution of the Anderson-Hubbard model under an interaction and disorder quench, where the Hubbard interaction and the disorder induced effective interaction are turned on instantaneously (at time $t=0$) within a system initially prepared in a non-interacting thermal equilibrium state at temperature $T=1/\beta = 0.025$. In other words, the interaction and the disorder are only present on the $\mathcal{C}_1\cup\mathcal{C}_2$ contour. We examine the momentum distribution function obtained from our real-time method. To standardize comparison to most previous studies which assumed disorder to be randomly distributed according to a box distribution given by $ P(V_i) = (1/2\Omega)\Theta(\Omega-|V_i|)$, we match the variance of the box and Gaussian distribution. Thus, in our numerical simulations the variance of the Gaussian distribution is given by $ W^2 = \Omega^2/3$.  The momentum distribution function $n(\varepsilon,t)$ is given by
    \begin{equation}
        n(\varepsilon_\mathbf{k},t) \equiv \expval{c^\dagger_{\mathbf{k},\sigma}(t)c_{\mathbf{k},\sigma}(t)} = -iG^{<}_{\mathbf{k},\sigma}(t,t) 
    \end{equation}
    where $\mathbf{k}$ and $\varepsilon_{\mathbf{k}}$ denote the momentum and energy dispersion of the electrons, respectively. Here,  $G^{<}_{\mathbf{k},\sigma}(t,t)$ is the lesser component of  the momentum resolved Green's function $G_{\mathbf{k},\sigma}(t,t')=-i\expval{\mathcal{T}_\mathcal{C}c_{\mathbf{k},\sigma}(t)c^\dagger_{\mathbf{k},\sigma}(t')}$. In Fig. \ref{fig:MomDistributionFunc}, we plot the momentum distribution function $n(\varepsilon,t)$ as a function of the band energy $\varepsilon$ and time $t$ for an interaction strength of $U=1.5$ in the clean system limit and with weak disorder, respectively . 

    From Figs. \hyperref[fig:MomDistributionFunc]{1(a)} and \hyperref[fig:MomDistributionFunc]{1(b)}, it is evident that $n(\varepsilon,t)$ evolves from a function with steep jump at $t=0$ to a smooth function at later times. In other words, the discontinuity $\Delta n(\varepsilon,t)$ at $\varepsilon=0$ smoothly decays to zero with time. Strictly speaking, the discontinuity in $n(\varepsilon=0,t)$ only exists for a non-interacting initial state at temperature $T=0$. However, in our work, we consider a temperature of $T=0.025$, where we observe a sharp jump near $\varepsilon=0$. This sharp change in $n(\varepsilon,t)$ at $\varepsilon=0$ is what we refer to as a discontinuity, although it should be noted that this is not a true mathematical discontinuity.  For an initial non-interacting state at half-filling, the discontinuity $\Delta n$ is given by \cite{EcksteinPRBInteractionquench_2010,EcksteinPRLThermalization_2009,ReviewNEDMFT}, 
    \begin{equation}
        \Delta n(\varepsilon=0,t) \equiv \Delta n (t) = |G^{\text{ret}}_{\varepsilon=0,\sigma}(t,0)|^2
    \end{equation}
    where, $G^{\text{ret}}_{\varepsilon_{\mathbf{k}},\sigma}(t,0) = -i\Theta(t)\expval{\{c_{\mathbf{k},\sigma}(0),c^\dagger_{\mathbf{k},\sigma}(t)\}}$ is the retarded component of the momentum resolved Green's function $G_{\mathbf{k},\sigma}(t,t')$.  Previous studies \cite{EcksteinPRLThermalization_2009,EcksteinPRBInteractionquench_2010} have demonstrated that $\Delta n(t)$ can be used to characterize the relaxation after the interaction quench in the Hubbard model. Here, we employ the Fermi surface discontinuity, $\Delta n(t)$ to characterize the relaxation dynamics of the system after the interaction and disorder quench.  

    \begin{figure}[htbp]
        \centering
        \begin{minipage}{\columnwidth}
            \centering
            \includegraphics[width=\columnwidth]{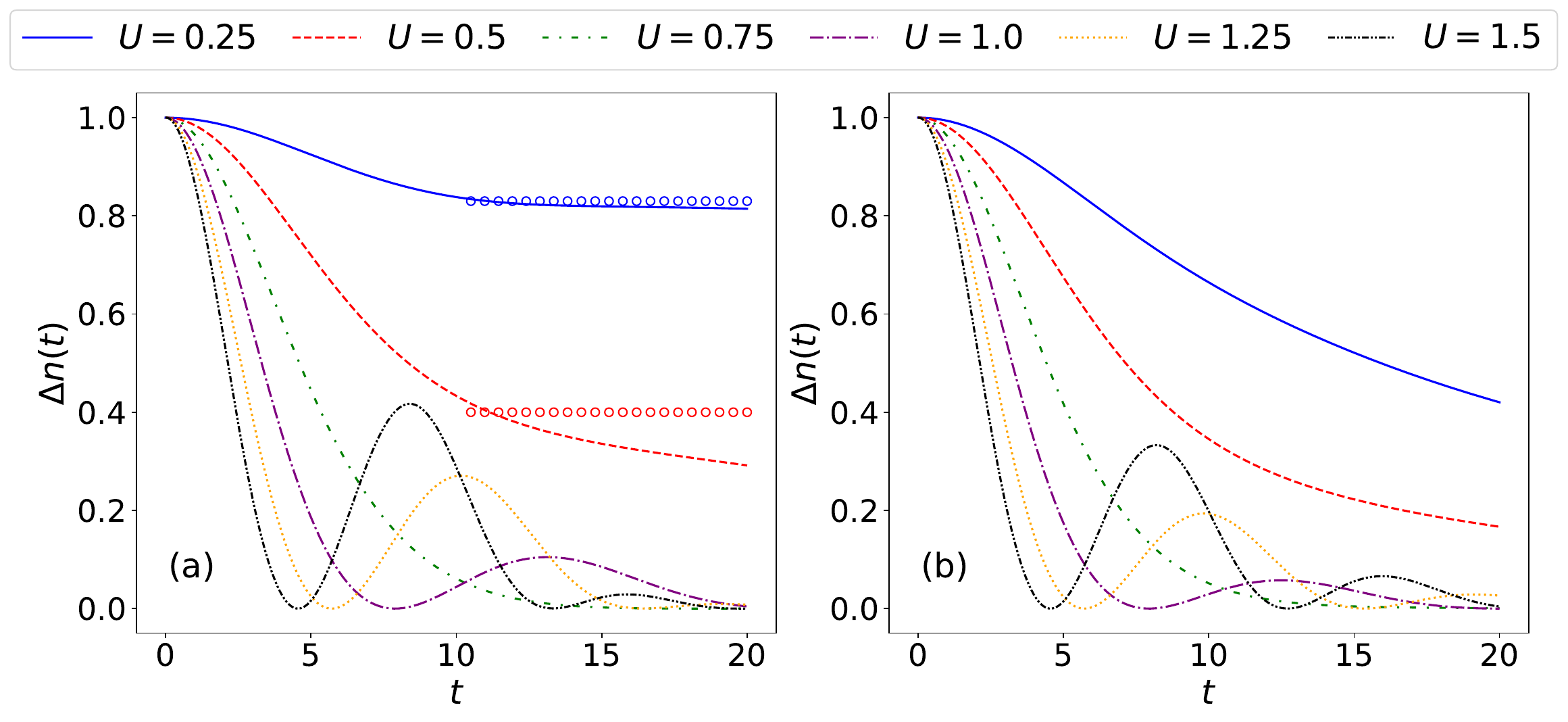}
        \end{minipage}
        \begin{minipage}{\columnwidth}
            \centering
            \includegraphics[width=\columnwidth]{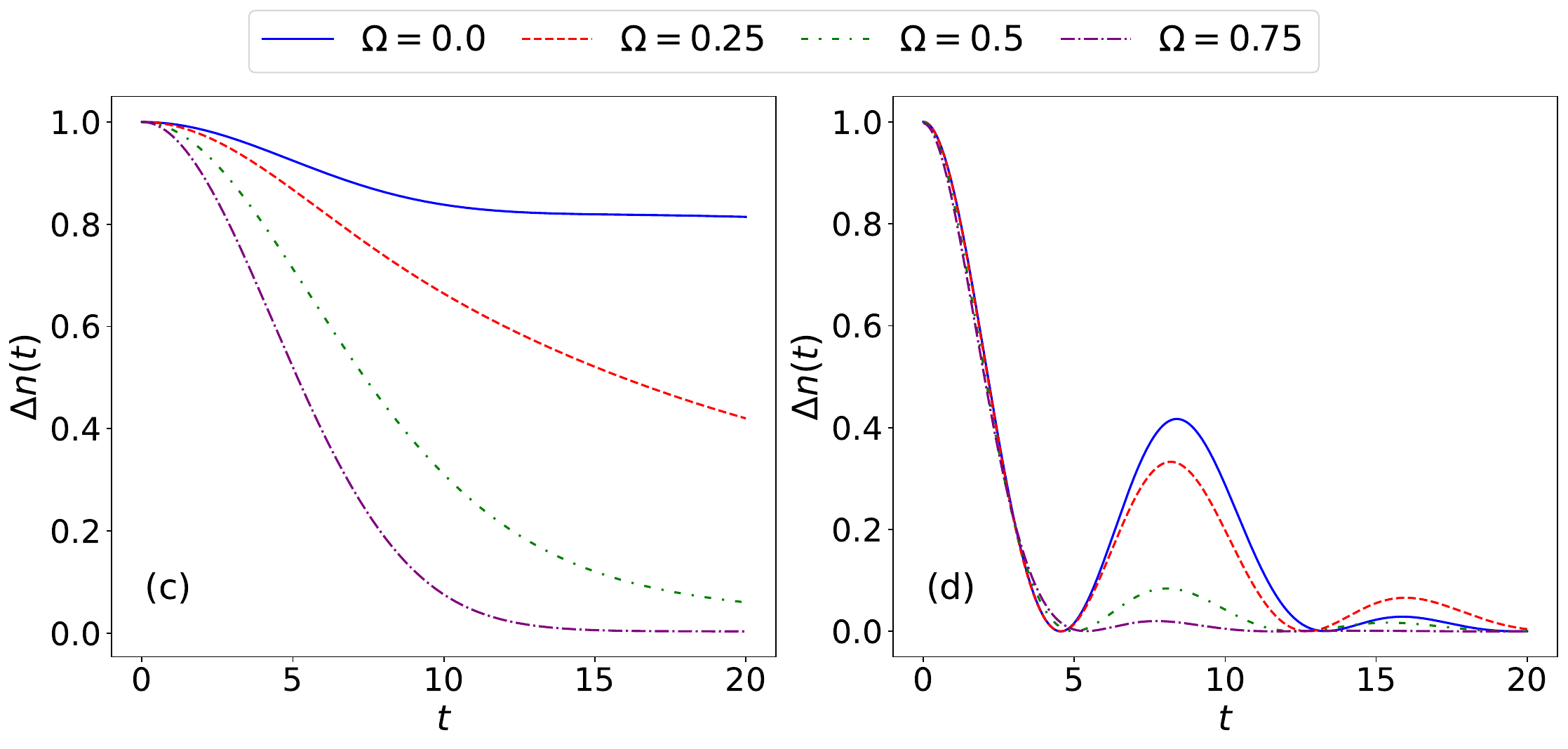}
        \end{minipage}
    \caption{Fermi surface discontinuity $\Delta n(t)$ as a function of time for the Anderson Hubbard model after interaction and disorder quenches to $U$ and $\Omega$, respectively. (a) clean system limit (b) weak disorder $\Omega=0.25$. Horizontal line (circles) in (a) denote the quasistationary value $\Delta n_{stat} = 2Z-1$ predicted in Ref. \cite{IntQuenchHubbard}, with $T=0$ quasiparticle weight Z taken from equilibrium DMFT data \cite{ZequibRef1,ZequibRef2}. Effect of disorder after an interaction quench to (c) $U=0.25$ and (d) $U=1.5$.}
        \label{fig:DeltaNEvst}
    \end{figure}
    
    In Fig. \hyperref[fig:DeltaNEvst]{2(a)}, we recover the results for the Hubbard model in the clean system limit. In particular, we observe that $\Delta n(t)$ behaves qualitatively different in the weak interaction and strong interaction regimes, separated by a sharp crossover at $U_c^{\text{dyn}} \approx 0.8$, in the units of the bandwidth. For quenches to $U < 0.75$, we observe that the value of $\Delta n(t)$ remains finite for short time scales whereas we observe collapse-and-revival oscillations around non-thermal values with approximate frequency of $2\pi/U$ for larges values of $U$. The observed behaviour at strong interaction is well understood in the atomic limit ($t_{hop}=0$), where the unitary evolution operator $e^{iHt}$ has a period of $2\pi/U$. In the weak interaction limit, we see that $\Delta n(t)$ attains quasistationary values for shorter timescales. The transient dynamics towards prethermalized plateaus to $\Delta n = 2Z-1$ for the metallic phase in the Hubbard model was predicted by Mockel and Kehrein \cite{IntQuenchHubbard} and numerically demonstrated in multiple studies \cite{EcksteinPRLThermalization_2009,PrethermalGGEKollar_2011,EcksteinPRBInteractionquench_2010}. Here, $Z$ is the quasiparticle weight in equilibrium at zero temperature. Around $U=U_c^{\text{dyn}}$, the system exhibits a rapid thermalization and a sharp crossover between two distinct transient dynamics. 

    We now introduce a disorder quench and analyze the resulting effects on the previously discussed phenomenon. In Fig. \hyperref[fig:DeltaNEvst]{2(b)}, we plot the Fermi surface discontinuity for the Anderson-Hubbard model for various interaction strengths and a weak value of disorder strength $\Omega=0.25$. Our results demonstrate that weak disorder tends to lower the quasistationary values of $\Delta n(t)$ for $U<U_c$ driving it towards a rapid thermalization regime. This is further illustrated by Fig. \hyperref[fig:DeltaNEvst]{2(c)}, which depicts $\Delta n(t)$ for an interaction quench to $U=0.25$ at varying values of disorder strength. As the disorder strength is increased from $\Omega=0.25$ to $\Omega=0.75$, $\Delta n(t)$ exhibits lower quasistationary values before reaching its thermal value. The influence of weak disorder can be understood through its effect on the quasiparticle weight $Z$, which governs the quasistationary values of the Fermi surface discontinuity $\Delta n(t)$. In the presence of weak disorder, enhanced scattering disrupts the coherence of quasiparticles, leading to a suppression of $Z$ \cite{Vollhardt_AndersonHubbard,Miranda_CPADMFT_2012June,ZimanyiDisorderInteractions}.  Alternatively, the effective interaction vertex arising from the disorder, as depicted in Eq. \eqref{eqn:eff_vertex}, contributes to an enhanced imaginary component of the self-energy, consequently suppressing the quasiparticle weight $Z$. In the strong interaction regime, the disorder tends to suppress the revival oscillations as evident from Fig. \hyperref[fig:DeltaNEvst]{2(d)}, where we show $\Delta n(t)$ for an interaction quench to $U=1.5$. We also notice that as the disorder strength increases, the oscillations vanish almost completely and $\Delta n(t)$ rapidly reaches a quasistationary value. 
    \begin{widetext}
        
    \begin{figure}[htbp]
        \centering
        \begin{minipage}{0.33\columnwidth}
            \centering
            \includegraphics[width=\columnwidth]{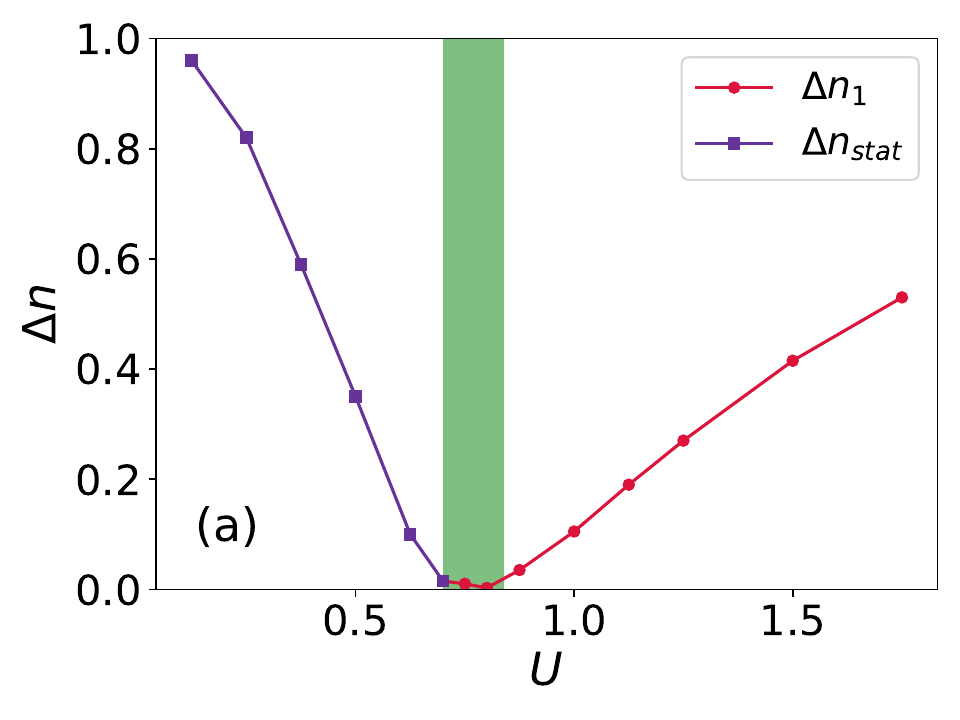} 
            \label{fig:DeltanvsUW0.0}
        \end{minipage}\hfill
        \begin{minipage}{0.33\columnwidth}
            \centering
            \includegraphics[width=\columnwidth]{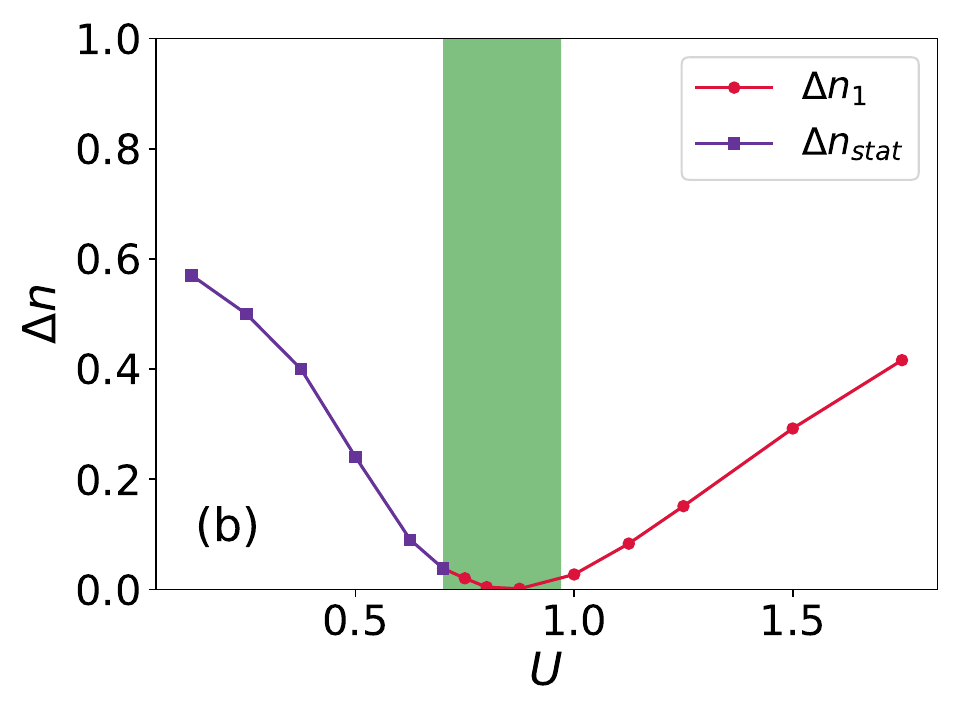} 
            \label{fig:DeltanvsUW0.25}
        \end{minipage}
        \begin{minipage}{0.33\columnwidth}
            \centering
        \includegraphics[width=\columnwidth]{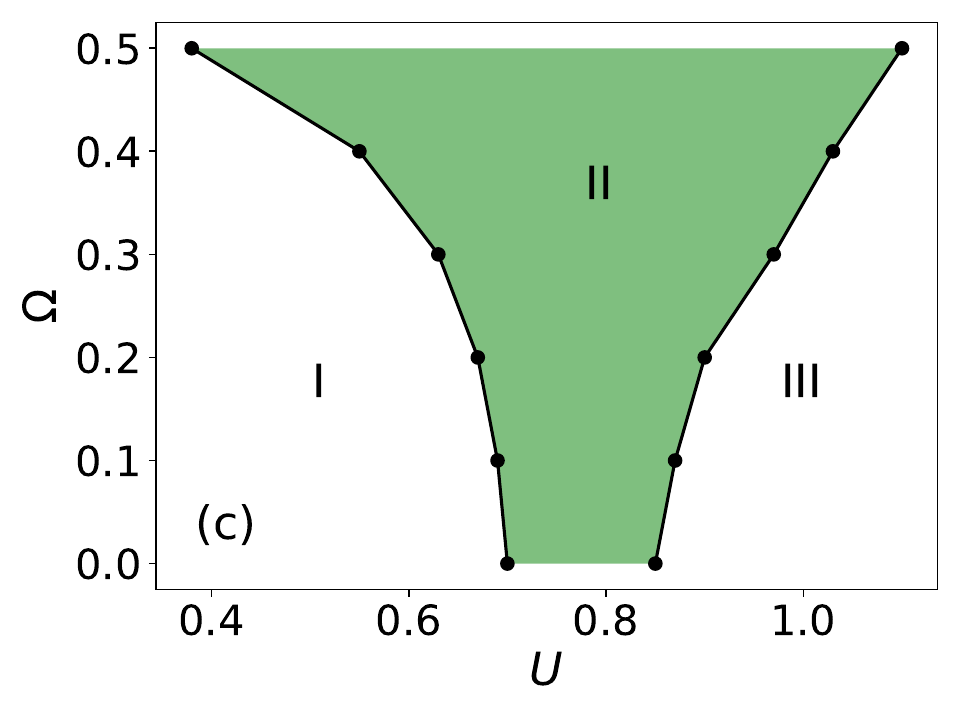}
        \label{fig:PhaseDiagram}
        \end{minipage}
        \caption{Prethermalization value of Fermi discontinuity $\Delta n_{stat}$ and first revival maximum $\Delta n_1$ as a function of $U$ for (a) clean system limit ($\Omega=0.0$) and (b) weak disorder ($\Omega=0.25$). The shaded region indicates the regime of fast thermalization. The behavior of $\Delta n$ in panel (a) agrees with the results for the Hubbard model \cite{EcksteinPRLThermalization_2009}. (c) The extended analysis of the various dynamical regimes observed in the non-equilibrium transient dynamics of the system over wider range of parameters in the $U-\Omega$ space. I and III denote prethermalization and collapse-and-revival oscillations regimes, respectively. II (shaded region) denotes the regime of fast thermalization.} 
        \label{fig:DeltaNvsU}
    \end{figure}
    
    \end{widetext}
    Fig.~\ref{fig:DeltaNvsU} depicts the crossover between the prethermalization regime and the collapse-and-revival oscillations regime, separated by a region of fast thermalization, as a function of the quenched interaction $U$. In the clean system limit (Fig. \hyperref[fig:DeltanvsUW0.0]{3(a)}), we observe a interaction strength, $U_c^{dyn} \approx 0.8$, separating the two regimes, consistent with previous reports \cite{EcksteinPRLThermalization_2009,EcksteinPRBInteractionquench_2010}. Fig. \hyperref[fig:DeltanvsUW0.25]{3(b)} shows that weak disorder broadens the region of fast thermalization (shaded area) likely due to the suppression of $Z$ as discussed previously.

    As illustrated in Fig. \hyperref[fig:DeltaNvsU]{3(c)}, we further characterize various dynamical regimes observed in the non-equilibrium transient dynamics of the system through a dynamical phase diagram as a function of $U$ and $\Omega$. The crossover between the disparate behaviors is hard to pinpoint, and the indicated regions in Fig. \hyperref[fig:DeltaNvsU]{3(c)} are only qualitative. Prior investigations of the clean Hubbard model ($\Omega=0$) have identified a Dynamical Phase Transition (DPT) separating regimes I and III \cite{DPT_review_2022,TDMFT_Quench2010,TDMFT_QuenchPRB}. However, our results suggests that disorder likely replaces this single DPT, inducing an extended crossover regime (II) where the system exhibits rapid dynamics for intermediate quenches. Similar behaviour of an extended crossover regime of rapid thermalization was reported in a DMFT study of an interaction quench in the Bose-Hubbard model \cite{BosonicDMFT}.  Additionally, studies suggest that introducing a finite doping away from half-filling in the Hubbard model washes out the DPT, resulting in a sharp crossover instead \cite{TDMFT_QuenchPRB,TDMFT_Quench2010}.

\textit{Conclusion} -- We introduce a formalism within the non-equilibrium Dynamical Mean Field Theory framework to account for the effects of static Gaussian random disorder. By leveraging the normalization property of the Schwinger-Keldysh partition function, we explicitly integrated over the distribution of disorder configurations, resulting in an effective time-dependent density-density interaction in addition to the Hubbard interaction. We use the formalism to investigate the real-time dynamics of the Anderson-Hubbard model following an interaction and disorder quench. We focused on the Fermi surface discontinuity as a key observable for relaxation. The method reproduced a behavior expected for the Hubbard model in the clean limit: the system exhibits prethermalization at weak interaction regime and collapse-and-revival oscillations in the strong interaction regimes, separated by a region of rapid thermalization. Strikingly, we observe that disorder promotes fast thermalization across a wider range of interaction strengths, contrasting with the strong disorder case, which is expected to slow down and perhaps even inhibit thermalization. The formalism presented here can be adapted to study a broad range of systems in which the competition and cooperation of Gaussian disorder and interaction control their physics, including the effects of disorder in broken symmetry phases such as superconductivity and magnetism. 

While the theoretical framework for disorder averaging we have presented here is exact within the context of DMFT, the practical implementation of our impurity solver introduces limitations. Specifically, bare perturbation theory (IPT) for nonequilibrium DMFT is inherently restricted to the weak-coupling regime. Developing a more robust impurity solver capable of handling a wider range of interactions, including time-dependent interactions, would significantly enhance the applicability of our nonequilibrium dynamical mean-field theory (DMFT) approach to disordered systems.  Finally, extending our analysis to the strong disorder regime is a crucial next step. By investigating the effects of strong disorder, we can potentially uncover Anderson localization transitions induced by the disorder, thereby deepening our understanding of the interplay between interactions and disorder in non-equilibrium systems. However, recovering Anderson localization within the local approximation of DMFT is notoriously difficult, even for systems in equilibrium. Some modifications of the present approach such as that in the typical medium theory may become necessary \cite{TMT_Dobrosavljevic,TMDCA_review}. Additionally, it can be adapted to the extended dynamical mean field theory, enabling the study of effects from spatially correlated randomness and long-range Coulomb interactions.

\textit{Acknowledgements} -- This manuscript is based on work supported by the US Department of Energy, Office of Science, Office of Basic Energy Sciences, under Award Number DE-SC0017861.  HFF is supported by the U.S. Department of Energy, Office of Science, Basic Energy Sciences, under Award number DE-SC0024139. HT is supported by the U.S. Department of Energy, Office of Science, Basic Energy Sciences, under Award number DE‐SC0024196 grant. This work used high-performance computational resources provided by the Louisiana Optical Network Initiative and HPC@LSU computing.
\appendix
\section{Appendix: Impurity solver and DMFT equations}
\begin{figure}[htbp]
    \centering
    \includegraphics[width=\columnwidth]{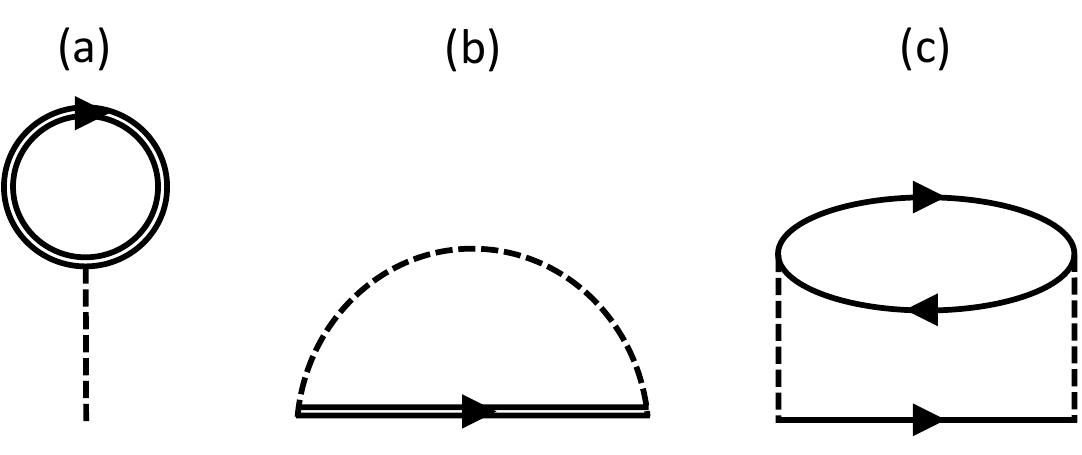}
    \caption{(a) Hartree and (b) Fock diagrams in the lowest order perturbation diagrams for the disorder vertex, $\bar{U}$. The double line in (a) and (b) represents in the interacting Green's function $G^{\bar{U}}_\sigma$. (c) The lowest order diagram for the Hubbard vertex, $U$. }
    \label{fig:Diagrams}
\end{figure}

In this section, we delve into the details of the DMFT equations and the impurity solver employed in our study. The DMFT algorithm begins with an initial guess for the hybridization function, $\Lambda_\sigma(t,t')$, which is used to calculate the non-interacting Green's function of the impurity site:
\begin{equation}
    \mathcal{G}^{-1}_{0,\sigma}(t,t')\equiv\Bigl(i\pdv{t} + \mu\Bigr)\delta_c(t,t') - \Lambda_\sigma(t,t')
\end{equation}
Subsequently, the self-energy of the impurity is computed using an impurity solver. In this study, we utilize perturbation theory to calculate the self-energy and Green's function of the impurity site. As stated in the main text, the bare perturbation (IPT) at the lowest order does not capture the expected physics of the interplay of disorder and interactions. To address this, we employ a self-consistent perturbation theory (SPT) for the disorder vertex,  
$\Bar{U}$, at the first-order perturbation diagram, while treating the Hubbard interaction, $U$, at the second-order bare IPT level. The lowest-order diagrams for the disorder vertex encompass Hartree and Fock contributions at the first order, as illustrated in Figs. \hyperref[fig:Diagrams]{4(a)} and \hyperref[fig:Diagrams]{4(b)}, respectively.

However, by imposing particle-hole symmetry through a fixed chemical potential, we eliminate all tadpole diagrams in Fig. \hyperref[fig:Diagrams]{4(a)}, leaving only the Fock diagram at the lowest order of the disorder vertex. This Fock diagram, represented by the double line in Fig. \hyperref[fig:Diagrams]{4(b)}, is calculated self-consistently within the perturbative approach using the interacting Green's function and is given by: 
\begin{equation}
    \Sigma^{\bar{U}}_\sigma (t,t') = -i\bar{U}G^{\bar{U}}_\sigma (t,t')
\end{equation}
This self-consistent perturbation theory for the disorder vertex is completed by the Dyson equation: 
\begin{equation}
    G^{\bar{U}}_\sigma (t,t') = (\mathcal{G}^{-1}_{0,\sigma} - \Sigma^{\bar{U}}_\sigma (t,t'))^{-1}
\end{equation}
The self-energy diagrams capturing the interplay of the disorder vertex and Hubbard interaction are determined by considering solely the lowest order diagram of the Hubbard vertex given by the self-energy diagram in Fig. \hyperref[fig:Diagrams]{4(c)}, computed using the self-consistently determined interacting Green's function, $G^{\bar{U}}_\sigma (t,t')$, rather than the bare Green's function:
\begin{equation}
    \Sigma^{U}_\sigma (t,t') = U(t)U(t')G^{\Bar{U}}_\sigma(t,t')G^{\Bar{U}}_\sigma(t,t')G^{\Bar{U}}_\sigma(t',t)
\end{equation}
The total self-energy of the impurity is thus given by: 
\begin{align}
        \Sigma_\sigma(t,t') &=  \Sigma^{U}_\sigma (t,t') + \Sigma^{\bar{U}}_\sigma (t,t')\\
        &= U(t)U(t')G^{\Bar{U}}_\sigma(t,t')G^{\Bar{U}}_\sigma(t,t')G^{\Bar{U}}_\sigma(t',t)\nonumber\\ 
        &\qquad- i\Bar{U}(t)G^{\Bar{U}}_\sigma(t,t'). \label{eqn:SelfEnergy}
    \end{align}
Further, the local Green's function is calculated from the impurity self-energy (Eq. \eqref{eqn:SelfEnergy}) using the Dyson equation:
\begin{equation}
    G_\sigma (t,t') = (\mathcal{G}^{-1}_{0,\sigma} - \Sigma_\sigma (t,t'))^{-1}
\end{equation}
For our model on the infinite dimensional Bethe lattice, the DMFT self-consistency condition reduces to:
\begin{equation}
    \Lambda_\sigma(t,t') = t^2_{hop}G_\sigma (t,t')
\end{equation}
This completes the description of the non-equilibrium DMFT algorithm as applied to the infinite-dimensional Bethe lattice.

\bibliography{references}

\begin{thebibliography}{68}%
\makeatletter
\providecommand \@ifxundefined [1]{%
 \@ifx{#1\undefined}
}%
\providecommand \@ifnum [1]{%
 \ifnum #1\expandafter \@firstoftwo
 \else \expandafter \@secondoftwo
 \fi
}%
\providecommand \@ifx [1]{%
 \ifx #1\expandafter \@firstoftwo
 \else \expandafter \@secondoftwo
 \fi
}%
\providecommand \natexlab [1]{#1}%
\providecommand \enquote  [1]{``#1''}%
\providecommand \bibnamefont  [1]{#1}%
\providecommand \bibfnamefont [1]{#1}%
\providecommand \citenamefont [1]{#1}%
\providecommand \href@noop [0]{\@secondoftwo}%
\providecommand \href [0]{\begingroup \@sanitize@url \@href}%
\providecommand \@href[1]{\@@startlink{#1}\@@href}%
\providecommand \@@href[1]{\endgroup#1\@@endlink}%
\providecommand \@sanitize@url [0]{\catcode `\\12\catcode `\$12\catcode `\&12\catcode `\#12\catcode `\^12\catcode `\_12\catcode `\%12\relax}%
\providecommand \@@startlink[1]{}%
\providecommand \@@endlink[0]{}%
\providecommand \url  [0]{\begingroup\@sanitize@url \@url }%
\providecommand \@url [1]{\endgroup\@href {#1}{\urlprefix }}%
\providecommand \urlprefix  [0]{URL }%
\providecommand \Eprint [0]{\href }%
\providecommand \doibase [0]{https://doi.org/}%
\providecommand \selectlanguage [0]{\@gobble}%
\providecommand \bibinfo  [0]{\@secondoftwo}%
\providecommand \bibfield  [0]{\@secondoftwo}%
\providecommand \translation [1]{[#1]}%
\providecommand \BibitemOpen [0]{}%
\providecommand \bibitemStop [0]{}%
\providecommand \bibitemNoStop [0]{.\EOS\space}%
\providecommand \EOS [0]{\spacefactor3000\relax}%
\providecommand \BibitemShut  [1]{\csname bibitem#1\endcsname}%
\let\auto@bib@innerbib\@empty
\bibitem [{\citenamefont {Greiner}\ \emph {et~al.}(2002)\citenamefont {Greiner}, \citenamefont {Mandel}, \citenamefont {Esslinger}, \citenamefont {H{\"a}nsch},\ and\ \citenamefont {Bloch}}]{Greiner2002}%
  \BibitemOpen
  \bibfield  {author} {\bibinfo {author} {\bibfnamefont {M.}~\bibnamefont {Greiner}}, \bibinfo {author} {\bibfnamefont {O.}~\bibnamefont {Mandel}}, \bibinfo {author} {\bibfnamefont {T.}~\bibnamefont {Esslinger}}, \bibinfo {author} {\bibfnamefont {T.~W.}\ \bibnamefont {H{\"a}nsch}},\ and\ \bibinfo {author} {\bibfnamefont {I.}~\bibnamefont {Bloch}},\ }\bibfield  {title} {\bibinfo {title} {Quantum phase transition from a superfluid to a mott insulator in a gas of ultracold atoms},\ }\href {https://doi.org/10.1038/415039a} {\bibfield  {journal} {\bibinfo  {journal} {Nature}\ }\textbf {\bibinfo {volume} {415}},\ \bibinfo {pages} {39} (\bibinfo {year} {2002})}\BibitemShut {NoStop}%
\bibitem [{\citenamefont {Eckstein}\ \emph {et~al.}(2009{\natexlab{a}})\citenamefont {Eckstein}, \citenamefont {Hackl}, \citenamefont {Kehrein}, \citenamefont {Kollar}, \citenamefont {Moeckel}, \citenamefont {Werner},\ and\ \citenamefont {Wolf}}]{Eckstein2009}%
  \BibitemOpen
  \bibfield  {author} {\bibinfo {author} {\bibfnamefont {M.}~\bibnamefont {Eckstein}}, \bibinfo {author} {\bibfnamefont {A.}~\bibnamefont {Hackl}}, \bibinfo {author} {\bibfnamefont {S.}~\bibnamefont {Kehrein}}, \bibinfo {author} {\bibfnamefont {M.}~\bibnamefont {Kollar}}, \bibinfo {author} {\bibfnamefont {M.}~\bibnamefont {Moeckel}}, \bibinfo {author} {\bibfnamefont {P.}~\bibnamefont {Werner}},\ and\ \bibinfo {author} {\bibfnamefont {F.~A.}\ \bibnamefont {Wolf}},\ }\bibfield  {title} {\bibinfo {title} {New theoretical approaches for correlated systems in nonequilibrium},\ }\href {https://doi.org/10.1140/epjst/e2010-01219-x} {\bibfield  {journal} {\bibinfo  {journal} {Eur. Phys. J. ST}\ }\textbf {\bibinfo {volume} {180}},\ \bibinfo {pages} {217} (\bibinfo {year} {2009}{\natexlab{a}})}\BibitemShut {NoStop}%
\bibitem [{\citenamefont {Wall}\ \emph {et~al.}(2011)\citenamefont {Wall}, \citenamefont {Brida}, \citenamefont {Clark}, \citenamefont {Ehrke}, \citenamefont {Jaksch}, \citenamefont {Ardavan}, \citenamefont {Bonora}, \citenamefont {Uemura}, \citenamefont {Takahashi}, \citenamefont {Hasegawa}, \citenamefont {Okamoto}, \citenamefont {Cerullo},\ and\ \citenamefont {Cavalleri}}]{Wall2011}%
  \BibitemOpen
  \bibfield  {author} {\bibinfo {author} {\bibfnamefont {S.}~\bibnamefont {Wall}}, \bibinfo {author} {\bibfnamefont {D.}~\bibnamefont {Brida}}, \bibinfo {author} {\bibfnamefont {S.~R.}\ \bibnamefont {Clark}}, \bibinfo {author} {\bibfnamefont {H.~P.}\ \bibnamefont {Ehrke}}, \bibinfo {author} {\bibfnamefont {D.}~\bibnamefont {Jaksch}}, \bibinfo {author} {\bibfnamefont {A.}~\bibnamefont {Ardavan}}, \bibinfo {author} {\bibfnamefont {S.}~\bibnamefont {Bonora}}, \bibinfo {author} {\bibfnamefont {H.}~\bibnamefont {Uemura}}, \bibinfo {author} {\bibfnamefont {Y.}~\bibnamefont {Takahashi}}, \bibinfo {author} {\bibfnamefont {T.}~\bibnamefont {Hasegawa}}, \bibinfo {author} {\bibfnamefont {H.}~\bibnamefont {Okamoto}}, \bibinfo {author} {\bibfnamefont {G.}~\bibnamefont {Cerullo}},\ and\ \bibinfo {author} {\bibfnamefont {A.}~\bibnamefont {Cavalleri}},\ }\bibfield  {title} {\bibinfo {title} {Quantum interference between charge excitation paths in a solid-state mott insulator},\ }\href {https://doi.org/10.1038/nphys1831}
  {\bibfield  {journal} {\bibinfo  {journal} {Nat. Phys.}\ }\textbf {\bibinfo {volume} {7}},\ \bibinfo {pages} {114} (\bibinfo {year} {2011})}\BibitemShut {NoStop}%
\bibitem [{\citenamefont {Eberlein}\ \emph {et~al.}(2017)\citenamefont {Eberlein}, \citenamefont {Kasper}, \citenamefont {Sachdev},\ and\ \citenamefont {Steinberg}}]{QuantumQuenchSYK}%
  \BibitemOpen
  \bibfield  {author} {\bibinfo {author} {\bibfnamefont {A.}~\bibnamefont {Eberlein}}, \bibinfo {author} {\bibfnamefont {V.}~\bibnamefont {Kasper}}, \bibinfo {author} {\bibfnamefont {S.}~\bibnamefont {Sachdev}},\ and\ \bibinfo {author} {\bibfnamefont {J.}~\bibnamefont {Steinberg}},\ }\bibfield  {title} {\bibinfo {title} {Quantum quench of the sachdev-ye-kitaev model},\ }\href {https://doi.org/10.1103/PhysRevB.96.205123} {\bibfield  {journal} {\bibinfo  {journal} {Phys. Rev. B}\ }\textbf {\bibinfo {volume} {96}},\ \bibinfo {pages} {205123} (\bibinfo {year} {2017})}\BibitemShut {NoStop}%
\bibitem [{\citenamefont {Aoki}\ \emph {et~al.}(2014)\citenamefont {Aoki}, \citenamefont {Tsuji}, \citenamefont {Eckstein}, \citenamefont {Kollar}, \citenamefont {Oka},\ and\ \citenamefont {Werner}}]{ReviewNEDMFT}%
  \BibitemOpen
  \bibfield  {author} {\bibinfo {author} {\bibfnamefont {H.}~\bibnamefont {Aoki}}, \bibinfo {author} {\bibfnamefont {N.}~\bibnamefont {Tsuji}}, \bibinfo {author} {\bibfnamefont {M.}~\bibnamefont {Eckstein}}, \bibinfo {author} {\bibfnamefont {M.}~\bibnamefont {Kollar}}, \bibinfo {author} {\bibfnamefont {T.}~\bibnamefont {Oka}},\ and\ \bibinfo {author} {\bibfnamefont {P.}~\bibnamefont {Werner}},\ }\bibfield  {title} {\bibinfo {title} {Nonequilibrium dynamical mean-field theory and its applications},\ }\href {https://doi.org/10.1103/RevModPhys.86.779} {\bibfield  {journal} {\bibinfo  {journal} {Rev. Mod. Phys.}\ }\textbf {\bibinfo {volume} {86}},\ \bibinfo {pages} {779} (\bibinfo {year} {2014})}\BibitemShut {NoStop}%
\bibitem [{\citenamefont {Lantz}\ \emph {et~al.}(2017)\citenamefont {Lantz}, \citenamefont {Mansart}, \citenamefont {Grieger}, \citenamefont {Boschetto}, \citenamefont {Nilforoushan}, \citenamefont {Papalazarou}, \citenamefont {Moisan}, \citenamefont {Perfetti}, \citenamefont {Jacques}, \citenamefont {Le~Bolloc'h}, \citenamefont {Laulh{\'e}}, \citenamefont {Ravy}, \citenamefont {Rueff}, \citenamefont {Glover}, \citenamefont {Hertlein}, \citenamefont {Hussain}, \citenamefont {Song}, \citenamefont {Chollet}, \citenamefont {Fabrizio},\ and\ \citenamefont {Marsi}}]{Lantz2017}%
  \BibitemOpen
  \bibfield  {author} {\bibinfo {author} {\bibfnamefont {G.}~\bibnamefont {Lantz}}, \bibinfo {author} {\bibfnamefont {B.}~\bibnamefont {Mansart}}, \bibinfo {author} {\bibfnamefont {D.}~\bibnamefont {Grieger}}, \bibinfo {author} {\bibfnamefont {D.}~\bibnamefont {Boschetto}}, \bibinfo {author} {\bibfnamefont {N.}~\bibnamefont {Nilforoushan}}, \bibinfo {author} {\bibfnamefont {E.}~\bibnamefont {Papalazarou}}, \bibinfo {author} {\bibfnamefont {N.}~\bibnamefont {Moisan}}, \bibinfo {author} {\bibfnamefont {L.}~\bibnamefont {Perfetti}}, \bibinfo {author} {\bibfnamefont {V.~L.~R.}\ \bibnamefont {Jacques}}, \bibinfo {author} {\bibfnamefont {D.}~\bibnamefont {Le~Bolloc'h}}, \bibinfo {author} {\bibfnamefont {C.}~\bibnamefont {Laulh{\'e}}}, \bibinfo {author} {\bibfnamefont {S.}~\bibnamefont {Ravy}}, \bibinfo {author} {\bibfnamefont {J.-P.}\ \bibnamefont {Rueff}}, \bibinfo {author} {\bibfnamefont {T.~E.}\ \bibnamefont {Glover}}, \bibinfo {author} {\bibfnamefont {M.~P.}\ \bibnamefont {Hertlein}}, \bibinfo {author}
  {\bibfnamefont {Z.}~\bibnamefont {Hussain}}, \bibinfo {author} {\bibfnamefont {S.}~\bibnamefont {Song}}, \bibinfo {author} {\bibfnamefont {M.}~\bibnamefont {Chollet}}, \bibinfo {author} {\bibfnamefont {M.}~\bibnamefont {Fabrizio}},\ and\ \bibinfo {author} {\bibfnamefont {M.}~\bibnamefont {Marsi}},\ }\bibfield  {title} {\bibinfo {title} {Ultrafast evolution and transient phases of a prototype out-of-equilibrium mott--hubbard material},\ }\href {https://doi.org/10.1038/ncomms13917} {\bibfield  {journal} {\bibinfo  {journal} {Nat. Commun.}\ }\textbf {\bibinfo {volume} {8}},\ \bibinfo {pages} {13917} (\bibinfo {year} {2017})}\BibitemShut {NoStop}%
\bibitem [{\citenamefont {Kollath}\ \emph {et~al.}(2007)\citenamefont {Kollath}, \citenamefont {L\"auchli},\ and\ \citenamefont {Altman}}]{QuenchBoseHubbard}%
  \BibitemOpen
  \bibfield  {author} {\bibinfo {author} {\bibfnamefont {C.}~\bibnamefont {Kollath}}, \bibinfo {author} {\bibfnamefont {A.~M.}\ \bibnamefont {L\"auchli}},\ and\ \bibinfo {author} {\bibfnamefont {E.}~\bibnamefont {Altman}},\ }\bibfield  {title} {\bibinfo {title} {Quench dynamics and nonequilibrium phase diagram of the bose-hubbard model},\ }\href {https://doi.org/10.1103/PhysRevLett.98.180601} {\bibfield  {journal} {\bibinfo  {journal} {Phys. Rev. Lett.}\ }\textbf {\bibinfo {volume} {98}},\ \bibinfo {pages} {180601} (\bibinfo {year} {2007})}\BibitemShut {NoStop}%
\bibitem [{\citenamefont {Barmettler}\ \emph {et~al.}(2009)\citenamefont {Barmettler}, \citenamefont {Punk}, \citenamefont {Gritsev}, \citenamefont {Demler},\ and\ \citenamefont {Altman}}]{Peter_HeisenbergChains}%
  \BibitemOpen
  \bibfield  {author} {\bibinfo {author} {\bibfnamefont {P.}~\bibnamefont {Barmettler}}, \bibinfo {author} {\bibfnamefont {M.}~\bibnamefont {Punk}}, \bibinfo {author} {\bibfnamefont {V.}~\bibnamefont {Gritsev}}, \bibinfo {author} {\bibfnamefont {E.}~\bibnamefont {Demler}},\ and\ \bibinfo {author} {\bibfnamefont {E.}~\bibnamefont {Altman}},\ }\bibfield  {title} {\bibinfo {title} {Relaxation of antiferromagnetic order in spin-$1/2$ chains following a quantum quench},\ }\href {https://doi.org/10.1103/PhysRevLett.102.130603} {\bibfield  {journal} {\bibinfo  {journal} {Phys. Rev. Lett.}\ }\textbf {\bibinfo {volume} {102}},\ \bibinfo {pages} {130603} (\bibinfo {year} {2009})}\BibitemShut {NoStop}%
\bibitem [{\citenamefont {Bhattacharya}\ \emph {et~al.}(2019)\citenamefont {Bhattacharya}, \citenamefont {Jatkar},\ and\ \citenamefont {Sorokhaibam}}]{ThermalizationSYK}%
  \BibitemOpen
  \bibfield  {author} {\bibinfo {author} {\bibfnamefont {R.}~\bibnamefont {Bhattacharya}}, \bibinfo {author} {\bibfnamefont {D.~P.}\ \bibnamefont {Jatkar}},\ and\ \bibinfo {author} {\bibfnamefont {N.}~\bibnamefont {Sorokhaibam}},\ }\bibfield  {title} {\bibinfo {title} {Quantum quenches and thermalization in syk models},\ }\href {https://doi.org/10.1007/JHEP07(2019)066} {\bibfield  {journal} {\bibinfo  {journal} {J. High Energy Phys.}\ }\textbf {\bibinfo {volume} {2019}}\bibinfo  {number} { (7)},\ \bibinfo {pages} {66}}\BibitemShut {NoStop}%
\bibitem [{\citenamefont {Moeckel}\ and\ \citenamefont {Kehrein}(2008)}]{IntQuenchHubbard}%
  \BibitemOpen
\bibfield  {number} {  }\bibfield  {author} {\bibinfo {author} {\bibfnamefont {M.}~\bibnamefont {Moeckel}}\ and\ \bibinfo {author} {\bibfnamefont {S.}~\bibnamefont {Kehrein}},\ }\bibfield  {title} {\bibinfo {title} {Interaction quench in the hubbard model},\ }\href {https://doi.org/10.1103/PhysRevLett.100.175702} {\bibfield  {journal} {\bibinfo  {journal} {Phys. Rev. Lett.}\ }\textbf {\bibinfo {volume} {100}},\ \bibinfo {pages} {175702} (\bibinfo {year} {2008})}\BibitemShut {NoStop}%
\bibitem [{\citenamefont {Eckstein}\ \emph {et~al.}(2009{\natexlab{b}})\citenamefont {Eckstein}, \citenamefont {Kollar},\ and\ \citenamefont {Werner}}]{EcksteinPRLThermalization_2009}%
  \BibitemOpen
  \bibfield  {author} {\bibinfo {author} {\bibfnamefont {M.}~\bibnamefont {Eckstein}}, \bibinfo {author} {\bibfnamefont {M.}~\bibnamefont {Kollar}},\ and\ \bibinfo {author} {\bibfnamefont {P.}~\bibnamefont {Werner}},\ }\bibfield  {title} {\bibinfo {title} {Thermalization after an interaction quench in the hubbard model},\ }\href {https://doi.org/10.1103/PhysRevLett.103.056403} {\bibfield  {journal} {\bibinfo  {journal} {Phys. Rev. Lett.}\ }\textbf {\bibinfo {volume} {103}},\ \bibinfo {pages} {056403} (\bibinfo {year} {2009}{\natexlab{b}})}\BibitemShut {NoStop}%
\bibitem [{\citenamefont {Reimann}\ and\ \citenamefont {Dabelow}(2019)}]{Peter_Typ_Prethermalization2019}%
  \BibitemOpen
  \bibfield  {author} {\bibinfo {author} {\bibfnamefont {P.}~\bibnamefont {Reimann}}\ and\ \bibinfo {author} {\bibfnamefont {L.}~\bibnamefont {Dabelow}},\ }\bibfield  {title} {\bibinfo {title} {Typicality of prethermalization},\ }\href {https://doi.org/10.1103/PhysRevLett.122.080603} {\bibfield  {journal} {\bibinfo  {journal} {Phys. Rev. Lett.}\ }\textbf {\bibinfo {volume} {122}},\ \bibinfo {pages} {080603} (\bibinfo {year} {2019})}\BibitemShut {NoStop}%
\bibitem [{\citenamefont {Balzer}\ \emph {et~al.}(2015)\citenamefont {Balzer}, \citenamefont {Wolf}, \citenamefont {McCulloch}, \citenamefont {Werner},\ and\ \citenamefont {Eckstein}}]{Karsten_PRX}%
  \BibitemOpen
  \bibfield  {author} {\bibinfo {author} {\bibfnamefont {K.}~\bibnamefont {Balzer}}, \bibinfo {author} {\bibfnamefont {F.~A.}\ \bibnamefont {Wolf}}, \bibinfo {author} {\bibfnamefont {I.~P.}\ \bibnamefont {McCulloch}}, \bibinfo {author} {\bibfnamefont {P.}~\bibnamefont {Werner}},\ and\ \bibinfo {author} {\bibfnamefont {M.}~\bibnamefont {Eckstein}},\ }\bibfield  {title} {\bibinfo {title} {Nonthermal melting of n\'eel order in the hubbard model},\ }\href {https://doi.org/10.1103/PhysRevX.5.031039} {\bibfield  {journal} {\bibinfo  {journal} {Phys. Rev. X}\ }\textbf {\bibinfo {volume} {5}},\ \bibinfo {pages} {031039} (\bibinfo {year} {2015})}\BibitemShut {NoStop}%
\bibitem [{\citenamefont {Marino}\ \emph {et~al.}(2022)\citenamefont {Marino}, \citenamefont {Eckstein}, \citenamefont {Foster},\ and\ \citenamefont {Rey}}]{DPT_review_2022}%
  \BibitemOpen
  \bibfield  {author} {\bibinfo {author} {\bibfnamefont {J.}~\bibnamefont {Marino}}, \bibinfo {author} {\bibfnamefont {M.}~\bibnamefont {Eckstein}}, \bibinfo {author} {\bibfnamefont {M.~S.}\ \bibnamefont {Foster}},\ and\ \bibinfo {author} {\bibfnamefont {A.~M.}\ \bibnamefont {Rey}},\ }\bibfield  {title} {\bibinfo {title} {Dynamical phase transitions in the collisionless pre-thermal states of isolated quantum systems: theory and experiments},\ }\href {https://doi.org/10.1088/1361-6633/ac906c} {\bibfield  {journal} {\bibinfo  {journal} {Rep. Prog. Phys.}\ }\textbf {\bibinfo {volume} {85}},\ \bibinfo {pages} {116001} (\bibinfo {year} {2022})}\BibitemShut {NoStop}%
\bibitem [{\citenamefont {Kollar}\ \emph {et~al.}(2011)\citenamefont {Kollar}, \citenamefont {Wolf},\ and\ \citenamefont {Eckstein}}]{PrethermalGGEKollar_2011}%
  \BibitemOpen
  \bibfield  {author} {\bibinfo {author} {\bibfnamefont {M.}~\bibnamefont {Kollar}}, \bibinfo {author} {\bibfnamefont {F.~A.}\ \bibnamefont {Wolf}},\ and\ \bibinfo {author} {\bibfnamefont {M.}~\bibnamefont {Eckstein}},\ }\bibfield  {title} {\bibinfo {title} {Generalized gibbs ensemble prediction of prethermalization plateaus and their relation to nonthermal steady states in integrable systems},\ }\href {https://doi.org/10.1103/PhysRevB.84.054304} {\bibfield  {journal} {\bibinfo  {journal} {Phys. Rev. B}\ }\textbf {\bibinfo {volume} {84}},\ \bibinfo {pages} {054304} (\bibinfo {year} {2011})}\BibitemShut {NoStop}%
\bibitem [{\citenamefont {Eckstein}\ \emph {et~al.}(2010)\citenamefont {Eckstein}, \citenamefont {Kollar},\ and\ \citenamefont {Werner}}]{EcksteinPRBInteractionquench_2010}%
  \BibitemOpen
  \bibfield  {author} {\bibinfo {author} {\bibfnamefont {M.}~\bibnamefont {Eckstein}}, \bibinfo {author} {\bibfnamefont {M.}~\bibnamefont {Kollar}},\ and\ \bibinfo {author} {\bibfnamefont {P.}~\bibnamefont {Werner}},\ }\bibfield  {title} {\bibinfo {title} {Interaction quench in the hubbard model: Relaxation of the spectral function and the optical conductivity},\ }\href {https://doi.org/10.1103/PhysRevB.81.115131} {\bibfield  {journal} {\bibinfo  {journal} {Phys. Rev. B}\ }\textbf {\bibinfo {volume} {81}},\ \bibinfo {pages} {115131} (\bibinfo {year} {2010})}\BibitemShut {NoStop}%
\bibitem [{\citenamefont {Schir\'o}\ and\ \citenamefont {Fabrizio}(2010)}]{TDMFT_Quench2010}%
  \BibitemOpen
  \bibfield  {author} {\bibinfo {author} {\bibfnamefont {M.}~\bibnamefont {Schir\'o}}\ and\ \bibinfo {author} {\bibfnamefont {M.}~\bibnamefont {Fabrizio}},\ }\bibfield  {title} {\bibinfo {title} {Time-dependent mean field theory for quench dynamics in correlated electron systems},\ }\href {https://doi.org/10.1103/PhysRevLett.105.076401} {\bibfield  {journal} {\bibinfo  {journal} {Phys. Rev. Lett.}\ }\textbf {\bibinfo {volume} {105}},\ \bibinfo {pages} {076401} (\bibinfo {year} {2010})}\BibitemShut {NoStop}%
\bibitem [{\citenamefont {Schir\'o}\ and\ \citenamefont {Fabrizio}(2011)}]{TDMFT_QuenchPRB}%
  \BibitemOpen
  \bibfield  {author} {\bibinfo {author} {\bibfnamefont {M.}~\bibnamefont {Schir\'o}}\ and\ \bibinfo {author} {\bibfnamefont {M.}~\bibnamefont {Fabrizio}},\ }\bibfield  {title} {\bibinfo {title} {Quantum quenches in the hubbard model: Time-dependent mean-field theory and the role of quantum fluctuations},\ }\href {https://doi.org/10.1103/PhysRevB.83.165105} {\bibfield  {journal} {\bibinfo  {journal} {Phys. Rev. B}\ }\textbf {\bibinfo {volume} {83}},\ \bibinfo {pages} {165105} (\bibinfo {year} {2011})}\BibitemShut {NoStop}%
\bibitem [{\citenamefont {Heyl}(2018)}]{DPT_review}%
  \BibitemOpen
  \bibfield  {author} {\bibinfo {author} {\bibfnamefont {M.}~\bibnamefont {Heyl}},\ }\bibfield  {title} {\bibinfo {title} {Dynamical quantum phase transitions: a review},\ }\href {https://doi.org/10.1088/1361-6633/aaaf9a} {\bibfield  {journal} {\bibinfo  {journal} {Rep. Prog. Phys.}\ }\textbf {\bibinfo {volume} {81}},\ \bibinfo {pages} {054001} (\bibinfo {year} {2018})}\BibitemShut {NoStop}%
\bibitem [{\citenamefont {Abanin}\ \emph {et~al.}(2019)\citenamefont {Abanin}, \citenamefont {Altman}, \citenamefont {Bloch},\ and\ \citenamefont {Serbyn}}]{MBL_RMP}%
  \BibitemOpen
  \bibfield  {author} {\bibinfo {author} {\bibfnamefont {D.~A.}\ \bibnamefont {Abanin}}, \bibinfo {author} {\bibfnamefont {E.}~\bibnamefont {Altman}}, \bibinfo {author} {\bibfnamefont {I.}~\bibnamefont {Bloch}},\ and\ \bibinfo {author} {\bibfnamefont {M.}~\bibnamefont {Serbyn}},\ }\bibfield  {title} {\bibinfo {title} {Colloquium: Many-body localization, thermalization, and entanglement},\ }\href {https://doi.org/10.1103/RevModPhys.91.021001} {\bibfield  {journal} {\bibinfo  {journal} {Rev. Mod. Phys.}\ }\textbf {\bibinfo {volume} {91}},\ \bibinfo {pages} {021001} (\bibinfo {year} {2019})}\BibitemShut {NoStop}%
\bibitem [{\citenamefont {Alet}\ and\ \citenamefont {Laflorencie}(2018)}]{Alet_Laflorencie_2018}%
  \BibitemOpen
  \bibfield  {author} {\bibinfo {author} {\bibfnamefont {F.}~\bibnamefont {Alet}}\ and\ \bibinfo {author} {\bibfnamefont {N.}~\bibnamefont {Laflorencie}},\ }\bibfield  {title} {\bibinfo {title} {Many-body localization: An introduction and selected topics},\ }\href {https://doi.org/https://doi.org/10.1016/j.crhy.2018.03.003} {\bibfield  {journal} {\bibinfo  {journal} {Comptes Rendus Physique}\ }\textbf {\bibinfo {volume} {19}},\ \bibinfo {pages} {498} (\bibinfo {year} {2018})}\BibitemShut {NoStop}%
\bibitem [{\citenamefont {Nandkishore}\ and\ \citenamefont {Huse}(2015)}]{Nandkishore_Huse_2015}%
  \BibitemOpen
  \bibfield  {author} {\bibinfo {author} {\bibfnamefont {R.}~\bibnamefont {Nandkishore}}\ and\ \bibinfo {author} {\bibfnamefont {D.~A.}\ \bibnamefont {Huse}},\ }\bibfield  {title} {\bibinfo {title} {Many-body localization and thermalization in quantum statistical mechanics},\ }\href {https://doi.org/https://doi.org/10.1146/annurev-conmatphys-031214-014726} {\bibfield  {journal} {\bibinfo  {journal} {Annu. Rev. Condens. Matter Phys.}\ }\textbf {\bibinfo {volume} {6}},\ \bibinfo {pages} {15} (\bibinfo {year} {2015})}\BibitemShut {NoStop}%
\bibitem [{\citenamefont {Vojta}(2019)}]{Vojta_disorder_review}%
  \BibitemOpen
  \bibfield  {author} {\bibinfo {author} {\bibfnamefont {T.}~\bibnamefont {Vojta}},\ }\bibfield  {title} {\bibinfo {title} {Disorder in quantum many-body systems},\ }\href {https://doi.org/https://doi.org/10.1146/annurev-conmatphys-031218-013433} {\bibfield  {journal} {\bibinfo  {journal} {Annu. Rev. Condens. Matter Phys.}\ }\textbf {\bibinfo {volume} {10}},\ \bibinfo {pages} {233} (\bibinfo {year} {2019})}\BibitemShut {NoStop}%
\bibitem [{\citenamefont {Elliott}\ \emph {et~al.}(1974)\citenamefont {Elliott}, \citenamefont {Krumhansl},\ and\ \citenamefont {Leath}}]{RevModPhys.46.465}%
  \BibitemOpen
  \bibfield  {author} {\bibinfo {author} {\bibfnamefont {R.~J.}\ \bibnamefont {Elliott}}, \bibinfo {author} {\bibfnamefont {J.~A.}\ \bibnamefont {Krumhansl}},\ and\ \bibinfo {author} {\bibfnamefont {P.~L.}\ \bibnamefont {Leath}},\ }\bibfield  {title} {\bibinfo {title} {The theory and properties of randomly disordered crystals and related physical systems},\ }\href {https://doi.org/10.1103/RevModPhys.46.465} {\bibfield  {journal} {\bibinfo  {journal} {Rev. Mod. Phys.}\ }\textbf {\bibinfo {volume} {46}},\ \bibinfo {pages} {465} (\bibinfo {year} {1974})}\BibitemShut {NoStop}%
\bibitem [{\citenamefont {Soven}(1967)}]{SovenP_CPA_Apr1967}%
  \BibitemOpen
  \bibfield  {author} {\bibinfo {author} {\bibfnamefont {P.}~\bibnamefont {Soven}},\ }\bibfield  {title} {\bibinfo {title} {Coherent-potential model of substitutional disordered alloys},\ }\href {https://doi.org/10.1103/PhysRev.156.809} {\bibfield  {journal} {\bibinfo  {journal} {Phys. Rev.}\ }\textbf {\bibinfo {volume} {156}},\ \bibinfo {pages} {809} (\bibinfo {year} {1967})}\BibitemShut {NoStop}%
\bibitem [{\citenamefont {Velick\'y}(1969)}]{Velicky_CPA_Aug1969}%
  \BibitemOpen
  \bibfield  {author} {\bibinfo {author} {\bibfnamefont {B.}~\bibnamefont {Velick\'y}},\ }\bibfield  {title} {\bibinfo {title} {Theory of electronic transport in disordered binary alloys: Coherent-potential approximation},\ }\href {https://doi.org/10.1103/PhysRev.184.614} {\bibfield  {journal} {\bibinfo  {journal} {Phys. Rev.}\ }\textbf {\bibinfo {volume} {184}},\ \bibinfo {pages} {614} (\bibinfo {year} {1969})}\BibitemShut {NoStop}%
\bibitem [{\citenamefont {Kirkpatrick}\ \emph {et~al.}(1970)\citenamefont {Kirkpatrick}, \citenamefont {Velick\'y},\ and\ \citenamefont {Ehrenreich}}]{Kirkpatrick_CPA_Apr1970}%
  \BibitemOpen
  \bibfield  {author} {\bibinfo {author} {\bibfnamefont {S.}~\bibnamefont {Kirkpatrick}}, \bibinfo {author} {\bibfnamefont {B.}~\bibnamefont {Velick\'y}},\ and\ \bibinfo {author} {\bibfnamefont {H.}~\bibnamefont {Ehrenreich}},\ }\bibfield  {title} {\bibinfo {title} {Paramagnetic $\mathrm{Ni}\mathrm{Cu}$ alloys: Electronic density of states in the coherent-potential approximation},\ }\href {https://doi.org/10.1103/PhysRevB.1.3250} {\bibfield  {journal} {\bibinfo  {journal} {Phys. Rev. B}\ }\textbf {\bibinfo {volume} {1}},\ \bibinfo {pages} {3250} (\bibinfo {year} {1970})}\BibitemShut {NoStop}%
\bibitem [{\citenamefont {Shiba}(1971)}]{Shiba_CPA_1971}%
  \BibitemOpen
  \bibfield  {author} {\bibinfo {author} {\bibfnamefont {H.}~\bibnamefont {Shiba}},\ }\bibfield  {title} {\bibinfo {title} {{A Reformulation of the Coherent Potential Approximation and Its Applications: }},\ }\href {https://doi.org/10.1143/PTP.46.77} {\bibfield  {journal} {\bibinfo  {journal} {Prog. Theor. Exp. Phys.}\ }\textbf {\bibinfo {volume} {46}},\ \bibinfo {pages} {77} (\bibinfo {year} {1971})}\BibitemShut {NoStop}%
\bibitem [{\citenamefont {Yonezawa}\ and\ \citenamefont {Morigaki}(1973)}]{Yonezawa1CPA_1973}%
  \BibitemOpen
  \bibfield  {author} {\bibinfo {author} {\bibfnamefont {F.}~\bibnamefont {Yonezawa}}\ and\ \bibinfo {author} {\bibfnamefont {K.}~\bibnamefont {Morigaki}},\ }\bibfield  {title} {\bibinfo {title} {Coherent potential approximation. basic concepts and applications},\ }\href {https://api.semanticscholar.org/CorpusID:122430096} {\bibfield  {journal} {\bibinfo  {journal} {Prog. theor. phys., Suppl.}\ }\textbf {\bibinfo {volume} {53}},\ \bibinfo {pages} {1} (\bibinfo {year} {1973})}\BibitemShut {NoStop}%
\bibitem [{\citenamefont {Metzner}\ and\ \citenamefont {Vollhardt}(1989)}]{WMetzner_DMFT_Jan1989}%
  \BibitemOpen
  \bibfield  {author} {\bibinfo {author} {\bibfnamefont {W.}~\bibnamefont {Metzner}}\ and\ \bibinfo {author} {\bibfnamefont {D.}~\bibnamefont {Vollhardt}},\ }\bibfield  {title} {\bibinfo {title} {Correlated lattice fermions in $d=\ensuremath{\infty}$ dimensions},\ }\href {https://doi.org/10.1103/PhysRevLett.62.324} {\bibfield  {journal} {\bibinfo  {journal} {Phys. Rev. Lett.}\ }\textbf {\bibinfo {volume} {62}},\ \bibinfo {pages} {324} (\bibinfo {year} {1989})}\BibitemShut {NoStop}%
\bibitem [{\citenamefont {Georges}\ \emph {et~al.}(1996)\citenamefont {Georges}, \citenamefont {Kotliar}, \citenamefont {Krauth},\ and\ \citenamefont {Rozenberg}}]{Georges_DMFTReview}%
  \BibitemOpen
  \bibfield  {author} {\bibinfo {author} {\bibfnamefont {A.}~\bibnamefont {Georges}}, \bibinfo {author} {\bibfnamefont {G.}~\bibnamefont {Kotliar}}, \bibinfo {author} {\bibfnamefont {W.}~\bibnamefont {Krauth}},\ and\ \bibinfo {author} {\bibfnamefont {M.~J.}\ \bibnamefont {Rozenberg}},\ }\bibfield  {title} {\bibinfo {title} {Dynamical mean-field theory of strongly correlated fermion systems and the limit of infinite dimensions},\ }\href {https://doi.org/10.1103/RevModPhys.68.13} {\bibfield  {journal} {\bibinfo  {journal} {Rev. Mod. Phys.}\ }\textbf {\bibinfo {volume} {68}},\ \bibinfo {pages} {13} (\bibinfo {year} {1996})}\BibitemShut {NoStop}%
\bibitem [{\citenamefont {Kotliar}\ \emph {et~al.}(2006)\citenamefont {Kotliar}, \citenamefont {Savrasov}, \citenamefont {Haule}, \citenamefont {Oudovenko}, \citenamefont {Parcollet},\ and\ \citenamefont {Marianetti}}]{Kotliar_DMFTReview}%
  \BibitemOpen
  \bibfield  {author} {\bibinfo {author} {\bibfnamefont {G.}~\bibnamefont {Kotliar}}, \bibinfo {author} {\bibfnamefont {S.~Y.}\ \bibnamefont {Savrasov}}, \bibinfo {author} {\bibfnamefont {K.}~\bibnamefont {Haule}}, \bibinfo {author} {\bibfnamefont {V.~S.}\ \bibnamefont {Oudovenko}}, \bibinfo {author} {\bibfnamefont {O.}~\bibnamefont {Parcollet}},\ and\ \bibinfo {author} {\bibfnamefont {C.~A.}\ \bibnamefont {Marianetti}},\ }\bibfield  {title} {\bibinfo {title} {Electronic structure calculations with dynamical mean-field theory},\ }\href {https://doi.org/10.1103/RevModPhys.78.865} {\bibfield  {journal} {\bibinfo  {journal} {Rev. Mod. Phys.}\ }\textbf {\bibinfo {volume} {78}},\ \bibinfo {pages} {865} (\bibinfo {year} {2006})}\BibitemShut {NoStop}%
\bibitem [{\citenamefont {Freericks}\ and\ \citenamefont {Zlati\ifmmode~\acute{c}\else \'{c}\fi{}}(2003)}]{Freericks_DMFTReview}%
  \BibitemOpen
  \bibfield  {author} {\bibinfo {author} {\bibfnamefont {J.~K.}\ \bibnamefont {Freericks}}\ and\ \bibinfo {author} {\bibfnamefont {V.}~\bibnamefont {Zlati\ifmmode~\acute{c}\else \'{c}\fi{}}},\ }\bibfield  {title} {\bibinfo {title} {Exact dynamical mean-field theory of the falicov-kimball model},\ }\href {https://doi.org/10.1103/RevModPhys.75.1333} {\bibfield  {journal} {\bibinfo  {journal} {Rev. Mod. Phys.}\ }\textbf {\bibinfo {volume} {75}},\ \bibinfo {pages} {1333} (\bibinfo {year} {2003})}\BibitemShut {NoStop}%
\bibitem [{\citenamefont {Maier}\ \emph {et~al.}(2005)\citenamefont {Maier}, \citenamefont {Jarrell}, \citenamefont {Pruschke},\ and\ \citenamefont {Hettler}}]{Thomas_ClusterDMFTReview}%
  \BibitemOpen
  \bibfield  {author} {\bibinfo {author} {\bibfnamefont {T.}~\bibnamefont {Maier}}, \bibinfo {author} {\bibfnamefont {M.}~\bibnamefont {Jarrell}}, \bibinfo {author} {\bibfnamefont {T.}~\bibnamefont {Pruschke}},\ and\ \bibinfo {author} {\bibfnamefont {M.~H.}\ \bibnamefont {Hettler}},\ }\bibfield  {title} {\bibinfo {title} {Quantum cluster theories},\ }\href {https://doi.org/10.1103/RevModPhys.77.1027} {\bibfield  {journal} {\bibinfo  {journal} {Rev. Mod. Phys.}\ }\textbf {\bibinfo {volume} {77}},\ \bibinfo {pages} {1027} (\bibinfo {year} {2005})}\BibitemShut {NoStop}%
\bibitem [{\citenamefont {Hettler}\ \emph {et~al.}(1998)\citenamefont {Hettler}, \citenamefont {Tahvildar-Zadeh}, \citenamefont {Jarrell}, \citenamefont {Pruschke},\ and\ \citenamefont {Krishnamurthy}}]{Hettler_ClusterDMFT_Sep1998}%
  \BibitemOpen
  \bibfield  {author} {\bibinfo {author} {\bibfnamefont {M.~H.}\ \bibnamefont {Hettler}}, \bibinfo {author} {\bibfnamefont {A.~N.}\ \bibnamefont {Tahvildar-Zadeh}}, \bibinfo {author} {\bibfnamefont {M.}~\bibnamefont {Jarrell}}, \bibinfo {author} {\bibfnamefont {T.}~\bibnamefont {Pruschke}},\ and\ \bibinfo {author} {\bibfnamefont {H.~R.}\ \bibnamefont {Krishnamurthy}},\ }\bibfield  {title} {\bibinfo {title} {Nonlocal dynamical correlations of strongly interacting electron systems},\ }\href {https://doi.org/10.1103/PhysRevB.58.R7475} {\bibfield  {journal} {\bibinfo  {journal} {Phys. Rev. B}\ }\textbf {\bibinfo {volume} {58}},\ \bibinfo {pages} {R7475} (\bibinfo {year} {1998})}\BibitemShut {NoStop}%
\bibitem [{\citenamefont {Hettler}\ \emph {et~al.}(2000)\citenamefont {Hettler}, \citenamefont {Mukherjee}, \citenamefont {Jarrell},\ and\ \citenamefont {Krishnamurthy}}]{Hettler_ClusterDMFT_May2000}%
  \BibitemOpen
  \bibfield  {author} {\bibinfo {author} {\bibfnamefont {M.~H.}\ \bibnamefont {Hettler}}, \bibinfo {author} {\bibfnamefont {M.}~\bibnamefont {Mukherjee}}, \bibinfo {author} {\bibfnamefont {M.}~\bibnamefont {Jarrell}},\ and\ \bibinfo {author} {\bibfnamefont {H.~R.}\ \bibnamefont {Krishnamurthy}},\ }\bibfield  {title} {\bibinfo {title} {Dynamical cluster approximation: Nonlocal dynamics of correlated electron systems},\ }\href {https://doi.org/10.1103/PhysRevB.61.12739} {\bibfield  {journal} {\bibinfo  {journal} {Phys. Rev. B}\ }\textbf {\bibinfo {volume} {61}},\ \bibinfo {pages} {12739} (\bibinfo {year} {2000})}\BibitemShut {NoStop}%
\bibitem [{\citenamefont {Kotliar}\ \emph {et~al.}(2001)\citenamefont {Kotliar}, \citenamefont {Savrasov}, \citenamefont {P\'alsson},\ and\ \citenamefont {Biroli}}]{Kotliar_CellularDMFT_Oct2001}%
  \BibitemOpen
  \bibfield  {author} {\bibinfo {author} {\bibfnamefont {G.}~\bibnamefont {Kotliar}}, \bibinfo {author} {\bibfnamefont {S.~Y.}\ \bibnamefont {Savrasov}}, \bibinfo {author} {\bibfnamefont {G.}~\bibnamefont {P\'alsson}},\ and\ \bibinfo {author} {\bibfnamefont {G.}~\bibnamefont {Biroli}},\ }\bibfield  {title} {\bibinfo {title} {Cellular dynamical mean field approach to strongly correlated systems},\ }\href {https://doi.org/10.1103/PhysRevLett.87.186401} {\bibfield  {journal} {\bibinfo  {journal} {Phys. Rev. Lett.}\ }\textbf {\bibinfo {volume} {87}},\ \bibinfo {pages} {186401} (\bibinfo {year} {2001})}\BibitemShut {NoStop}%
\bibitem [{\citenamefont {Jarrell}\ \emph {et~al.}(2001)\citenamefont {Jarrell}, \citenamefont {Maier}, \citenamefont {Huscroft},\ and\ \citenamefont {Moukouri}}]{Jarrell_ClusterDMFT_Oct2001}%
  \BibitemOpen
  \bibfield  {author} {\bibinfo {author} {\bibfnamefont {M.}~\bibnamefont {Jarrell}}, \bibinfo {author} {\bibfnamefont {T.}~\bibnamefont {Maier}}, \bibinfo {author} {\bibfnamefont {C.}~\bibnamefont {Huscroft}},\ and\ \bibinfo {author} {\bibfnamefont {S.}~\bibnamefont {Moukouri}},\ }\bibfield  {title} {\bibinfo {title} {Quantum monte carlo algorithm for nonlocal corrections to the dynamical mean-field approximation},\ }\href {https://doi.org/10.1103/PhysRevB.64.195130} {\bibfield  {journal} {\bibinfo  {journal} {Phys. Rev. B}\ }\textbf {\bibinfo {volume} {64}},\ \bibinfo {pages} {195130} (\bibinfo {year} {2001})}\BibitemShut {NoStop}%
\bibitem [{\citenamefont {Terletska}\ \emph {et~al.}(2018)\citenamefont {Terletska}, \citenamefont {Zhang}, \citenamefont {Tam}, \citenamefont {Berlijn}, \citenamefont {Chioncel}, \citenamefont {Vidhyadhiraja},\ and\ \citenamefont {Jarrell}}]{TMDCA_review}%
  \BibitemOpen
  \bibfield  {author} {\bibinfo {author} {\bibfnamefont {H.}~\bibnamefont {Terletska}}, \bibinfo {author} {\bibfnamefont {Y.}~\bibnamefont {Zhang}}, \bibinfo {author} {\bibfnamefont {K.-M.}\ \bibnamefont {Tam}}, \bibinfo {author} {\bibfnamefont {T.}~\bibnamefont {Berlijn}}, \bibinfo {author} {\bibfnamefont {L.}~\bibnamefont {Chioncel}}, \bibinfo {author} {\bibfnamefont {N.}~\bibnamefont {Vidhyadhiraja}},\ and\ \bibinfo {author} {\bibfnamefont {M.}~\bibnamefont {Jarrell}},\ }\bibfield  {title} {\bibinfo {title} {Systematic quantum cluster typical medium method for the study of localization in strongly disordered electronic systems},\ }\href {https://doi.org/10.3390/app8122401} {\bibfield  {journal} {\bibinfo  {journal} {Appl. Sci.}\ }\textbf {\bibinfo {volume} {8}},\ \bibinfo {pages} {2401} (\bibinfo {year} {2018})}\BibitemShut {NoStop}%
\bibitem [{\citenamefont {Tam}\ \emph {et~al.}(2021)\citenamefont {Tam}, \citenamefont {Terletska}, \citenamefont {Berlijn}, \citenamefont {Chioncel},\ and\ \citenamefont {Moreno}}]{realspaceTMT}%
  \BibitemOpen
  \bibfield  {author} {\bibinfo {author} {\bibfnamefont {K.-M.}\ \bibnamefont {Tam}}, \bibinfo {author} {\bibfnamefont {H.}~\bibnamefont {Terletska}}, \bibinfo {author} {\bibfnamefont {T.}~\bibnamefont {Berlijn}}, \bibinfo {author} {\bibfnamefont {L.}~\bibnamefont {Chioncel}},\ and\ \bibinfo {author} {\bibfnamefont {J.}~\bibnamefont {Moreno}},\ }\bibfield  {title} {\bibinfo {title} {Real space quantum cluster formulation for the typical medium theory of anderson localization},\ }\href {https://doi.org/10.3390/cryst11111282} {\bibfield  {journal} {\bibinfo  {journal} {Crystals}\ }\textbf {\bibinfo {volume} {11}},\ \bibinfo {pages} {1282} (\bibinfo {year} {2021})}\BibitemShut {NoStop}%
\bibitem [{\citenamefont {Fotso}\ \emph {et~al.}(2012)\citenamefont {Fotso}, \citenamefont {Yang}, \citenamefont {Chen}, \citenamefont {Pathak}, \citenamefont {Moreno}, \citenamefont {Jarrell}, \citenamefont {Mikelsons}, \citenamefont {Khatami},\ and\ \citenamefont {Galanakis}}]{Fotso2012}%
  \BibitemOpen
  \bibfield  {author} {\bibinfo {author} {\bibfnamefont {H.}~\bibnamefont {Fotso}}, \bibinfo {author} {\bibfnamefont {S.}~\bibnamefont {Yang}}, \bibinfo {author} {\bibfnamefont {K.}~\bibnamefont {Chen}}, \bibinfo {author} {\bibfnamefont {S.}~\bibnamefont {Pathak}}, \bibinfo {author} {\bibfnamefont {J.}~\bibnamefont {Moreno}}, \bibinfo {author} {\bibfnamefont {M.}~\bibnamefont {Jarrell}}, \bibinfo {author} {\bibfnamefont {K.}~\bibnamefont {Mikelsons}}, \bibinfo {author} {\bibfnamefont {E.}~\bibnamefont {Khatami}},\ and\ \bibinfo {author} {\bibfnamefont {D.}~\bibnamefont {Galanakis}},\ }\bibinfo {title} {Dynamical cluster approximation},\ in\ \href {https://doi.org/10.1007/978-3-642-21831-6_9} {\emph {\bibinfo {booktitle} {Strongly Correlated Systems: Theoretical Methods}}},\ \bibinfo {editor} {edited by\ \bibinfo {editor} {\bibfnamefont {A.}~\bibnamefont {Avella}}\ and\ \bibinfo {editor} {\bibfnamefont {F.}~\bibnamefont {Mancini}}}\ (\bibinfo  {publisher} {Springer Berlin Heidelberg},\ \bibinfo {address}
  {Berlin, Heidelberg},\ \bibinfo {year} {2012})\ pp.\ \bibinfo {pages} {271--302}\BibitemShut {NoStop}%
\bibitem [{\citenamefont {Fotso}\ \emph {et~al.}(2022)\citenamefont {Fotso}, \citenamefont {Tam},\ and\ \citenamefont {Moreno}}]{Fotso_2022}%
  \BibitemOpen
  \bibfield  {author} {\bibinfo {author} {\bibfnamefont {H.~F.}\ \bibnamefont {Fotso}}, \bibinfo {author} {\bibfnamefont {K.-M.}\ \bibnamefont {Tam}},\ and\ \bibinfo {author} {\bibfnamefont {J.}~\bibnamefont {Moreno}},\ }\bibfield  {title} {\bibinfo {title} {Beyond quantum cluster theories: multiscale approaches for strongly correlated systems},\ }\href {https://doi.org/10.1088/2058-9565/ac676b} {\bibfield  {journal} {\bibinfo  {journal} {Quantum Sci. Technol.}\ }\textbf {\bibinfo {volume} {7}},\ \bibinfo {pages} {033001} (\bibinfo {year} {2022})}\BibitemShut {NoStop}%
\bibitem [{\citenamefont {Janis\ifmmode~\check{}\else \v{}\fi{}}\ and\ \citenamefont {Vollhardt}(1992)}]{Janis_CPADMFT_1992Dec}%
  \BibitemOpen
  \bibfield  {author} {\bibinfo {author} {\bibfnamefont {V.}~\bibnamefont {Janis\ifmmode~\check{}\else \v{}\fi{}}}\ and\ \bibinfo {author} {\bibfnamefont {D.}~\bibnamefont {Vollhardt}},\ }\bibfield  {title} {\bibinfo {title} {Coupling of quantum degrees of freedom in strongly interacting disordered electron systems},\ }\href {https://doi.org/10.1103/PhysRevB.46.15712} {\bibfield  {journal} {\bibinfo  {journal} {Phys. Rev. B}\ }\textbf {\bibinfo {volume} {46}},\ \bibinfo {pages} {15712} (\bibinfo {year} {1992})}\BibitemShut {NoStop}%
\bibitem [{\citenamefont {Kuchinskii}\ \emph {et~al.}(2010)\citenamefont {Kuchinskii}, \citenamefont {Kuleeva}, \citenamefont {Nekrasov},\ and\ \citenamefont {Sadovskii}}]{Kuchinskii_CPADMFT_2010}%
  \BibitemOpen
  \bibfield  {author} {\bibinfo {author} {\bibfnamefont {E.~Z.}\ \bibnamefont {Kuchinskii}}, \bibinfo {author} {\bibfnamefont {N.~A.}\ \bibnamefont {Kuleeva}}, \bibinfo {author} {\bibfnamefont {I.~A.}\ \bibnamefont {Nekrasov}},\ and\ \bibinfo {author} {\bibfnamefont {M.~V.}\ \bibnamefont {Sadovskii}},\ }\bibfield  {title} {\bibinfo {title} {Two-dimensional anderson-hubbard model in the dmft + $\sigma$ approximation},\ }\href {https://doi.org/10.1134/S1063776110020160} {\bibfield  {journal} {\bibinfo  {journal} {J. Exp. Theor. Phys.}\ }\textbf {\bibinfo {volume} {110}},\ \bibinfo {pages} {325} (\bibinfo {year} {2010})}\BibitemShut {NoStop}%
\bibitem [{\citenamefont {Semmler}\ \emph {et~al.}(2011)\citenamefont {Semmler}, \citenamefont {Byczuk},\ and\ \citenamefont {Hofstetter}}]{Semmler_CPADMFT_2011Sept}%
  \BibitemOpen
  \bibfield  {author} {\bibinfo {author} {\bibfnamefont {D.}~\bibnamefont {Semmler}}, \bibinfo {author} {\bibfnamefont {K.}~\bibnamefont {Byczuk}},\ and\ \bibinfo {author} {\bibfnamefont {W.}~\bibnamefont {Hofstetter}},\ }\bibfield  {title} {\bibinfo {title} {Anderson-hubbard model with box disorder: Statistical dynamical mean-field theory investigation},\ }\href {https://doi.org/10.1103/PhysRevB.84.115113} {\bibfield  {journal} {\bibinfo  {journal} {Phys. Rev. B}\ }\textbf {\bibinfo {volume} {84}},\ \bibinfo {pages} {115113} (\bibinfo {year} {2011})}\BibitemShut {NoStop}%
\bibitem [{\citenamefont {Miranda}\ and\ \citenamefont {Dobrosavljević}(2012)}]{Miranda_CPADMFT_2012June}%
  \BibitemOpen
  \bibfield  {author} {\bibinfo {author} {\bibfnamefont {E.}~\bibnamefont {Miranda}}\ and\ \bibinfo {author} {\bibfnamefont {V.}~\bibnamefont {Dobrosavljević}},\ }\bibfield  {title} {\bibinfo {title} {{1616 Dynamical Mean-field Theories of Correlation and Disorder}},\ }in\ \href {https://doi.org/10.1093/acprof:oso/9780199592593.003.0006} {\emph {\bibinfo {booktitle} {{Conductor-Insulator Quantum Phase Transitions}}}}\ (\bibinfo  {publisher} {Oxford University Press},\ \bibinfo {year} {2012})\BibitemShut {NoStop}%
\bibitem [{\citenamefont {Weh}\ \emph {et~al.}(2021)\citenamefont {Weh}, \citenamefont {Zhang}, \citenamefont {\"Ostlin}, \citenamefont {Terletska}, \citenamefont {Bauernfeind}, \citenamefont {Tam}, \citenamefont {Evertz}, \citenamefont {Byczuk}, \citenamefont {Vollhardt},\ and\ \citenamefont {Chioncel}}]{Weh_CPADMFT_2021Jul}%
  \BibitemOpen
  \bibfield  {author} {\bibinfo {author} {\bibfnamefont {A.}~\bibnamefont {Weh}}, \bibinfo {author} {\bibfnamefont {Y.}~\bibnamefont {Zhang}}, \bibinfo {author} {\bibfnamefont {A.}~\bibnamefont {\"Ostlin}}, \bibinfo {author} {\bibfnamefont {H.}~\bibnamefont {Terletska}}, \bibinfo {author} {\bibfnamefont {D.}~\bibnamefont {Bauernfeind}}, \bibinfo {author} {\bibfnamefont {K.-M.}\ \bibnamefont {Tam}}, \bibinfo {author} {\bibfnamefont {H.~G.}\ \bibnamefont {Evertz}}, \bibinfo {author} {\bibfnamefont {K.}~\bibnamefont {Byczuk}}, \bibinfo {author} {\bibfnamefont {D.}~\bibnamefont {Vollhardt}},\ and\ \bibinfo {author} {\bibfnamefont {L.}~\bibnamefont {Chioncel}},\ }\bibfield  {title} {\bibinfo {title} {Dynamical mean-field theory of the anderson-hubbard model with local and nonlocal disorder in tensor formulation},\ }\href {https://doi.org/10.1103/PhysRevB.104.045127} {\bibfield  {journal} {\bibinfo  {journal} {Phys. Rev. B}\ }\textbf {\bibinfo {volume} {104}},\ \bibinfo {pages} {045127} (\bibinfo {year}
  {2021})}\BibitemShut {NoStop}%
\bibitem [{\citenamefont {Ekuma}\ \emph {et~al.}(2015)\citenamefont {Ekuma}, \citenamefont {Yang}, \citenamefont {Terletska}, \citenamefont {Tam}, \citenamefont {Vidhyadhiraja}, \citenamefont {Moreno},\ and\ \citenamefont {Jarrell}}]{Ekuma_etal_2015}%
  \BibitemOpen
  \bibfield  {author} {\bibinfo {author} {\bibfnamefont {C.~E.}\ \bibnamefont {Ekuma}}, \bibinfo {author} {\bibfnamefont {S.-X.}\ \bibnamefont {Yang}}, \bibinfo {author} {\bibfnamefont {H.}~\bibnamefont {Terletska}}, \bibinfo {author} {\bibfnamefont {K.-M.}\ \bibnamefont {Tam}}, \bibinfo {author} {\bibfnamefont {N.~S.}\ \bibnamefont {Vidhyadhiraja}}, \bibinfo {author} {\bibfnamefont {J.}~\bibnamefont {Moreno}},\ and\ \bibinfo {author} {\bibfnamefont {M.}~\bibnamefont {Jarrell}},\ }\bibfield  {title} {\bibinfo {title} {Metal-insulator transition in a weakly interacting disordered electron system},\ }\href {https://doi.org/10.1103/PhysRevB.92.201114} {\bibfield  {journal} {\bibinfo  {journal} {Phys. Rev. B}\ }\textbf {\bibinfo {volume} {92}},\ \bibinfo {pages} {201114} (\bibinfo {year} {2015})}\BibitemShut {NoStop}%
\bibitem [{\citenamefont {Dong}\ and\ \citenamefont {Cheng}(2017)}]{dong2017pump}%
  \BibitemOpen
  \bibfield  {author} {\bibinfo {author} {\bibfnamefont {P.-T.}\ \bibnamefont {Dong}}\ and\ \bibinfo {author} {\bibfnamefont {J.-X.}\ \bibnamefont {Cheng}},\ }\bibfield  {title} {\bibinfo {title} {Pump-probe microscopy: theory, instrumentation, and applications},\ }\href {https://www.spectroscopyonline.com/view/vol-32-no-4-spectroscopy-april-2017-regular-issue-pdf} {\bibfield  {journal} {\bibinfo  {journal} {Spectroscopy}\ }\textbf {\bibinfo {volume} {32}},\ \bibinfo {pages} {24} (\bibinfo {year} {2017})}\BibitemShut {NoStop}%
\bibitem [{\citenamefont {Gr{\"u}nbein}\ \emph {et~al.}(2020)\citenamefont {Gr{\"u}nbein}, \citenamefont {Stricker}, \citenamefont {Nass~Kovacs}, \citenamefont {Kloos}, \citenamefont {Doak}, \citenamefont {Shoeman}, \citenamefont {Reinstein}, \citenamefont {Lecler}, \citenamefont {Haacke},\ and\ \citenamefont {Schlichting}}]{grunbein2020illumination}%
  \BibitemOpen
  \bibfield  {author} {\bibinfo {author} {\bibfnamefont {M.~L.}\ \bibnamefont {Gr{\"u}nbein}}, \bibinfo {author} {\bibfnamefont {M.}~\bibnamefont {Stricker}}, \bibinfo {author} {\bibfnamefont {G.}~\bibnamefont {Nass~Kovacs}}, \bibinfo {author} {\bibfnamefont {M.}~\bibnamefont {Kloos}}, \bibinfo {author} {\bibfnamefont {R.~B.}\ \bibnamefont {Doak}}, \bibinfo {author} {\bibfnamefont {R.~L.}\ \bibnamefont {Shoeman}}, \bibinfo {author} {\bibfnamefont {J.}~\bibnamefont {Reinstein}}, \bibinfo {author} {\bibfnamefont {S.}~\bibnamefont {Lecler}}, \bibinfo {author} {\bibfnamefont {S.}~\bibnamefont {Haacke}},\ and\ \bibinfo {author} {\bibfnamefont {I.}~\bibnamefont {Schlichting}},\ }\bibfield  {title} {\bibinfo {title} {Illumination guidelines for ultrafast pump--probe experiments by serial femtosecond crystallography},\ }\href {https://doi.org/10.1038/s41592-020-0847-3} {\bibfield  {journal} {\bibinfo  {journal} {Nat. Methods}\ }\textbf {\bibinfo {volume} {17}},\ \bibinfo {pages} {681} (\bibinfo {year}
  {2020})}\BibitemShut {NoStop}%
\bibitem [{\citenamefont {Fischer}\ \emph {et~al.}(2016)\citenamefont {Fischer}, \citenamefont {Wilson}, \citenamefont {Robles},\ and\ \citenamefont {Warren}}]{fischer2016invited}%
  \BibitemOpen
  \bibfield  {author} {\bibinfo {author} {\bibfnamefont {M.~C.}\ \bibnamefont {Fischer}}, \bibinfo {author} {\bibfnamefont {J.~W.}\ \bibnamefont {Wilson}}, \bibinfo {author} {\bibfnamefont {F.~E.}\ \bibnamefont {Robles}},\ and\ \bibinfo {author} {\bibfnamefont {W.~S.}\ \bibnamefont {Warren}},\ }\bibfield  {title} {\bibinfo {title} {{Invited Review Article: Pump-probe microscopy}},\ }\href {https://doi.org/10.1063/1.4943211} {\bibfield  {journal} {\bibinfo  {journal} {Rev. Sci. Instrum.}\ }\textbf {\bibinfo {volume} {87}},\ \bibinfo {pages} {031101} (\bibinfo {year} {2016})}\BibitemShut {NoStop}%
\bibitem [{\citenamefont {Dohner}\ \emph {et~al.}(2022)\citenamefont {Dohner}, \citenamefont {Terletska}, \citenamefont {Tam}, \citenamefont {Moreno},\ and\ \citenamefont {Fotso}}]{Dohner_NEDMFTCPA}%
  \BibitemOpen
  \bibfield  {author} {\bibinfo {author} {\bibfnamefont {E.}~\bibnamefont {Dohner}}, \bibinfo {author} {\bibfnamefont {H.}~\bibnamefont {Terletska}}, \bibinfo {author} {\bibfnamefont {K.-M.}\ \bibnamefont {Tam}}, \bibinfo {author} {\bibfnamefont {J.}~\bibnamefont {Moreno}},\ and\ \bibinfo {author} {\bibfnamefont {H.~F.}\ \bibnamefont {Fotso}},\ }\bibfield  {title} {\bibinfo {title} {Nonequilibrium $\text{DMFT}+\text{CPA}$ for correlated disordered systems},\ }\href {https://doi.org/10.1103/PhysRevB.106.195156} {\bibfield  {journal} {\bibinfo  {journal} {Phys. Rev. B}\ }\textbf {\bibinfo {volume} {106}},\ \bibinfo {pages} {195156} (\bibinfo {year} {2022})}\BibitemShut {NoStop}%
\bibitem [{\citenamefont {Yan}\ and\ \citenamefont {Werner}(2023)}]{YanJiawei_NEDMFTCPA}%
  \BibitemOpen
  \bibfield  {author} {\bibinfo {author} {\bibfnamefont {J.}~\bibnamefont {Yan}}\ and\ \bibinfo {author} {\bibfnamefont {P.}~\bibnamefont {Werner}},\ }\bibfield  {title} {\bibinfo {title} {Dynamical mean-field approach to disordered interacting systems and applications to the quantum transport problem},\ }\href {https://doi.org/10.1103/PhysRevB.108.125143} {\bibfield  {journal} {\bibinfo  {journal} {Phys. Rev. B}\ }\textbf {\bibinfo {volume} {108}},\ \bibinfo {pages} {125143} (\bibinfo {year} {2023})}\BibitemShut {NoStop}%
\bibitem [{\citenamefont {Dohner}\ \emph {et~al.}(2023)\citenamefont {Dohner}, \citenamefont {Terletska},\ and\ \citenamefont {Fotso}}]{Dohner_Thermalization_Oct2023}%
  \BibitemOpen
  \bibfield  {author} {\bibinfo {author} {\bibfnamefont {E.}~\bibnamefont {Dohner}}, \bibinfo {author} {\bibfnamefont {H.}~\bibnamefont {Terletska}},\ and\ \bibinfo {author} {\bibfnamefont {H.~F.}\ \bibnamefont {Fotso}},\ }\bibfield  {title} {\bibinfo {title} {Thermalization of a disordered interacting system under an interaction quench},\ }\href {https://doi.org/10.1103/PhysRevB.108.144202} {\bibfield  {journal} {\bibinfo  {journal} {Phys. Rev. B}\ }\textbf {\bibinfo {volume} {108}},\ \bibinfo {pages} {144202} (\bibinfo {year} {2023})}\BibitemShut {NoStop}%
\bibitem [{\citenamefont {Rangi}\ \emph {et~al.}(2024)\citenamefont {Rangi}, \citenamefont {Moreno},\ and\ \citenamefont {Tam}}]{RangiOTOC_2024}%
  \BibitemOpen
  \bibfield  {author} {\bibinfo {author} {\bibfnamefont {C.}~\bibnamefont {Rangi}}, \bibinfo {author} {\bibfnamefont {J.}~\bibnamefont {Moreno}},\ and\ \bibinfo {author} {\bibfnamefont {K.-M.}\ \bibnamefont {Tam}},\ }\bibfield  {title} {\bibinfo {title} {Out of time order correlation of the hubbard model with random local disorder},\ }\href {https://doi.org/10.1063/5.0206420} {\bibfield  {journal} {\bibinfo  {journal} {Chaos: An Interdisciplinary Journal of Nonlinear Science}\ }\textbf {\bibinfo {volume} {34}},\ \bibinfo {pages} {073143} (\bibinfo {year} {2024})}\BibitemShut {NoStop}%
\bibitem [{\citenamefont {Horbach}\ and\ \citenamefont {Schön}(1993)}]{Horbach_Schon_1993}%
  \BibitemOpen
  \bibfield  {author} {\bibinfo {author} {\bibfnamefont {M.~L.}\ \bibnamefont {Horbach}}\ and\ \bibinfo {author} {\bibfnamefont {G.}~\bibnamefont {Schön}},\ }\bibfield  {title} {\bibinfo {title} {Dynamic nonlinear σ-model of electron localization},\ }\href {https://doi.org/https://doi.org/10.1002/andp.19935050106} {\bibfield  {journal} {\bibinfo  {journal} {Annalen der Physik}\ }\textbf {\bibinfo {volume} {505}},\ \bibinfo {pages} {51} (\bibinfo {year} {1993})}\BibitemShut {NoStop}%
\bibitem [{\citenamefont {Kamenev}\ and\ \citenamefont {Andreev}(1999)}]{Kamenev_Andreev_1999}%
  \BibitemOpen
  \bibfield  {author} {\bibinfo {author} {\bibfnamefont {A.}~\bibnamefont {Kamenev}}\ and\ \bibinfo {author} {\bibfnamefont {A.}~\bibnamefont {Andreev}},\ }\bibfield  {title} {\bibinfo {title} {Electron-electron interactions in disordered metals: Keldysh formalism},\ }\href {https://doi.org/10.1103/PhysRevB.60.2218} {\bibfield  {journal} {\bibinfo  {journal} {Phys. Rev. B}\ }\textbf {\bibinfo {volume} {60}},\ \bibinfo {pages} {2218} (\bibinfo {year} {1999})}\BibitemShut {NoStop}%
\bibitem [{\citenamefont {Chamon}\ \emph {et~al.}(1999)\citenamefont {Chamon}, \citenamefont {Ludwig},\ and\ \citenamefont {Nayak}}]{CChamon1999Jul}%
  \BibitemOpen
  \bibfield  {author} {\bibinfo {author} {\bibfnamefont {C.}~\bibnamefont {Chamon}}, \bibinfo {author} {\bibfnamefont {A.~W.~W.}\ \bibnamefont {Ludwig}},\ and\ \bibinfo {author} {\bibfnamefont {C.}~\bibnamefont {Nayak}},\ }\bibfield  {title} {\bibinfo {title} {Schwinger-keldysh approach to disordered and interacting electron systems: Derivation of finkelstein's renormalization-group equations},\ }\href {https://doi.org/10.1103/PhysRevB.60.2239} {\bibfield  {journal} {\bibinfo  {journal} {Phys. Rev. B}\ }\textbf {\bibinfo {volume} {60}},\ \bibinfo {pages} {2239} (\bibinfo {year} {1999})}\BibitemShut {NoStop}%
\bibitem [{\citenamefont {Landsman}\ and\ \citenamefont {{van Weert}}(1987)}]{landsman_1987}%
  \BibitemOpen
  \bibfield  {author} {\bibinfo {author} {\bibfnamefont {N.}~\bibnamefont {Landsman}}\ and\ \bibinfo {author} {\bibfnamefont {C.}~\bibnamefont {{van Weert}}},\ }\bibfield  {title} {\bibinfo {title} {Real- and imaginary-time field theory at finite temperature and density},\ }\href {https://doi.org/https://doi.org/10.1016/0370-1573(87)90121-9} {\bibfield  {journal} {\bibinfo  {journal} {Phys. Rep.}\ }\textbf {\bibinfo {volume} {145}},\ \bibinfo {pages} {141} (\bibinfo {year} {1987})}\BibitemShut {NoStop}%
\bibitem [{\citenamefont {Freericks}(2008)}]{Freericks_BlochOsc_Feb2008}%
  \BibitemOpen
  \bibfield  {author} {\bibinfo {author} {\bibfnamefont {J.~K.}\ \bibnamefont {Freericks}},\ }\bibfield  {title} {\bibinfo {title} {Quenching bloch oscillations in a strongly correlated material: Nonequilibrium dynamical mean-field theory},\ }\href {https://doi.org/10.1103/PhysRevB.77.075109} {\bibfield  {journal} {\bibinfo  {journal} {Phys. Rev. B}\ }\textbf {\bibinfo {volume} {77}},\ \bibinfo {pages} {075109} (\bibinfo {year} {2008})}\BibitemShut {NoStop}%
\bibitem [{\citenamefont {Tsuji}\ and\ \citenamefont {Werner}(2013)}]{WeakCoupling_Naoto}%
  \BibitemOpen
  \bibfield  {author} {\bibinfo {author} {\bibfnamefont {N.}~\bibnamefont {Tsuji}}\ and\ \bibinfo {author} {\bibfnamefont {P.}~\bibnamefont {Werner}},\ }\bibfield  {title} {\bibinfo {title} {Nonequilibrium dynamical mean-field theory based on weak-coupling perturbation expansions: Application to dynamical symmetry breaking in the hubbard model},\ }\href {https://doi.org/10.1103/PhysRevB.88.165115} {\bibfield  {journal} {\bibinfo  {journal} {Phys. Rev. B}\ }\textbf {\bibinfo {volume} {88}},\ \bibinfo {pages} {165115} (\bibinfo {year} {2013})}\BibitemShut {NoStop}%
\bibitem [{\citenamefont {Schüler}\ \emph {et~al.}(2020)\citenamefont {Schüler}, \citenamefont {Golež}, \citenamefont {Murakami}, \citenamefont {Bittner}, \citenamefont {Herrmann}, \citenamefont {Strand}, \citenamefont {Werner},\ and\ \citenamefont {Eckstein}}]{NESSi}%
  \BibitemOpen
  \bibfield  {author} {\bibinfo {author} {\bibfnamefont {M.}~\bibnamefont {Schüler}}, \bibinfo {author} {\bibfnamefont {D.}~\bibnamefont {Golež}}, \bibinfo {author} {\bibfnamefont {Y.}~\bibnamefont {Murakami}}, \bibinfo {author} {\bibfnamefont {N.}~\bibnamefont {Bittner}}, \bibinfo {author} {\bibfnamefont {A.}~\bibnamefont {Herrmann}}, \bibinfo {author} {\bibfnamefont {H.~U.}\ \bibnamefont {Strand}}, \bibinfo {author} {\bibfnamefont {P.}~\bibnamefont {Werner}},\ and\ \bibinfo {author} {\bibfnamefont {M.}~\bibnamefont {Eckstein}},\ }\bibfield  {title} {\bibinfo {title} {Nessi: The non-equilibrium systems simulation package},\ }\href {https://doi.org/https://doi.org/10.1016/j.cpc.2020.107484} {\bibfield  {journal} {\bibinfo  {journal} {Computer Physics Communications}\ }\textbf {\bibinfo {volume} {257}},\ \bibinfo {pages} {107484} (\bibinfo {year} {2020})}\BibitemShut {NoStop}%
\bibitem [{\citenamefont {Bulla}(1999)}]{ZequibRef1}%
  \BibitemOpen
  \bibfield  {author} {\bibinfo {author} {\bibfnamefont {R.}~\bibnamefont {Bulla}},\ }\bibfield  {title} {\bibinfo {title} {Zero temperature metal-insulator transition in the infinite-dimensional hubbard model},\ }\href {https://doi.org/10.1103/PhysRevLett.83.136} {\bibfield  {journal} {\bibinfo  {journal} {Phys. Rev. Lett.}\ }\textbf {\bibinfo {volume} {83}},\ \bibinfo {pages} {136} (\bibinfo {year} {1999})}\BibitemShut {NoStop}%
\bibitem [{\citenamefont {Noack}\ and\ \citenamefont {Gebhard}(1999)}]{ZequibRef2}%
  \BibitemOpen
  \bibfield  {author} {\bibinfo {author} {\bibfnamefont {R.~M.}\ \bibnamefont {Noack}}\ and\ \bibinfo {author} {\bibfnamefont {F.}~\bibnamefont {Gebhard}},\ }\bibfield  {title} {\bibinfo {title} {Mott-hubbard transition in infinite dimensions},\ }\href {https://doi.org/10.1103/PhysRevLett.82.1915} {\bibfield  {journal} {\bibinfo  {journal} {Phys. Rev. Lett.}\ }\textbf {\bibinfo {volume} {82}},\ \bibinfo {pages} {1915} (\bibinfo {year} {1999})}\BibitemShut {NoStop}%
\bibitem [{\citenamefont {Byczuk}\ \emph {et~al.}(2005)\citenamefont {Byczuk}, \citenamefont {Hofstetter},\ and\ \citenamefont {Vollhardt}}]{Vollhardt_AndersonHubbard}%
  \BibitemOpen
  \bibfield  {author} {\bibinfo {author} {\bibfnamefont {K.}~\bibnamefont {Byczuk}}, \bibinfo {author} {\bibfnamefont {W.}~\bibnamefont {Hofstetter}},\ and\ \bibinfo {author} {\bibfnamefont {D.}~\bibnamefont {Vollhardt}},\ }\bibfield  {title} {\bibinfo {title} {Mott-hubbard transition versus anderson localization in correlated electron systems with disorder},\ }\href {https://doi.org/10.1103/PhysRevLett.94.056404} {\bibfield  {journal} {\bibinfo  {journal} {Phys. Rev. Lett.}\ }\textbf {\bibinfo {volume} {94}},\ \bibinfo {pages} {056404} (\bibinfo {year} {2005})}\BibitemShut {NoStop}%
\bibitem [{\citenamefont {Zimanyi}\ and\ \citenamefont {Abrahams}(1990)}]{ZimanyiDisorderInteractions}%
  \BibitemOpen
  \bibfield  {author} {\bibinfo {author} {\bibfnamefont {G.~T.}\ \bibnamefont {Zimanyi}}\ and\ \bibinfo {author} {\bibfnamefont {E.}~\bibnamefont {Abrahams}},\ }\bibfield  {title} {\bibinfo {title} {Disorder and interactions in the hubbard model},\ }\href {https://doi.org/10.1103/PhysRevLett.64.2719} {\bibfield  {journal} {\bibinfo  {journal} {Phys. Rev. Lett.}\ }\textbf {\bibinfo {volume} {64}},\ \bibinfo {pages} {2719} (\bibinfo {year} {1990})}\BibitemShut {NoStop}%
\bibitem [{\citenamefont {Strand}\ \emph {et~al.}(2015)\citenamefont {Strand}, \citenamefont {Eckstein},\ and\ \citenamefont {Werner}}]{BosonicDMFT}%
  \BibitemOpen
  \bibfield  {author} {\bibinfo {author} {\bibfnamefont {H.~U.~R.}\ \bibnamefont {Strand}}, \bibinfo {author} {\bibfnamefont {M.}~\bibnamefont {Eckstein}},\ and\ \bibinfo {author} {\bibfnamefont {P.}~\bibnamefont {Werner}},\ }\bibfield  {title} {\bibinfo {title} {Nonequilibrium dynamical mean-field theory for bosonic lattice models},\ }\href {https://doi.org/10.1103/PhysRevX.5.011038} {\bibfield  {journal} {\bibinfo  {journal} {Phys. Rev. X}\ }\textbf {\bibinfo {volume} {5}},\ \bibinfo {pages} {011038} (\bibinfo {year} {2015})}\BibitemShut {NoStop}%
\bibitem [{\citenamefont {Dobrosavljevi{\'c}}\ \emph {et~al.}(2003)\citenamefont {Dobrosavljevi{\'c}}, \citenamefont {Pastor},\ and\ \citenamefont {Nikoli{\'c}}}]{TMT_Dobrosavljevic}%
  \BibitemOpen
  \bibfield  {author} {\bibinfo {author} {\bibfnamefont {V.}~\bibnamefont {Dobrosavljevi{\'c}}}, \bibinfo {author} {\bibfnamefont {A.}~\bibnamefont {Pastor}},\ and\ \bibinfo {author} {\bibfnamefont {B.}~\bibnamefont {Nikoli{\'c}}},\ }\bibfield  {title} {\bibinfo {title} {Typical medium theory of anderson localization: A local order parameter approach to strong-disorder effects},\ }\href {https://doi.org/10.1209/epl/i2003-00364-5} {\bibfield  {journal} {\bibinfo  {journal} {EPL}\ }\textbf {\bibinfo {volume} {62}},\ \bibinfo {pages} {76} (\bibinfo {year} {2003})}\BibitemShut {NoStop}%
\end{thebibliography}%
\newpage
\appendix
\onecolumngrid
\end{document}